\documentclass{article}
 \pdfoutput=1

\usepackage{arxiv}
\usepackage[utf8]{inputenc} % allow utf-8 input
\usepackage[T1]{fontenc}    % use 8-bit T1 fonts
\usepackage{hyperref}       % hyperlinks
\usepackage{url}            % simple URL typesetting
\usepackage{booktabs}       % professional-quality tables
\usepackage{amsfonts}       % blackboard math symbols
\usepackage{nicefrac}       % compact symbols for 1/2, etc.
\usepackage{doi}            % doi
\usepackage{fancyhdr}       % header
\usepackage{graphicx}       % graphics

\usepackage{float}
\usepackage{multirow}
\usepackage{array}
\usepackage{bigstrut}
\usepackage{longtable}
\usepackage{makecell}
\usepackage{layouts}

\graphicspath{{./figures/}} 

\title{Abnormal Functional Brain Network Connectivity Associated with Alzheimer’s Disease}

\author{ 
\href{https://orcid.org/0000-0003-2754-3649}{\includegraphics[scale=0.06]{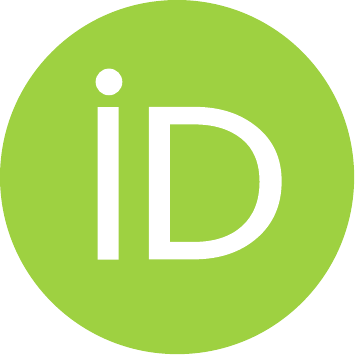}\hspace{1mm} Yongcheng Yao} \\
Department of Imaging and Interventional Radiology \\
The Chinese University of Hong Kong \\
Hong Kong\\
\texttt{yao\_yongchneng@link.cuhk.edu.hk} \\
}

\begin{document}
\maketitle

\begin{abstract}
The study's objective is to explore the distinctions in the functional brain network connectivity between Alzheimer's Disease (AD) patients and normal controls using Functional Magnetic Resonance Imaging (fMRI). The study included 590 individuals, with 175 having AD dementia and 415 age-, gender-, and handedness-matched normal controls. The connectivity of functional brain networks was measured using ROI-to-ROI and ROI-to-Voxel connectivity analyses. The findings reveal a general decrease in functional connectivity among the AD group in comparison to the normal control group. These results advance our comprehension of AD pathophysiology and could assist in identifying AD biomarkers.
\end{abstract}

% keywords can be removed

\section{Introduction}
Alzheimer's Disease (AD) is a chronic neurodegenerative disease that primarily affects the elderly, characterized by cognitive decline, language problems, memory disturbances (especially short-term memory), and disorientation. As the disease progresses, severe bodily dysfunction and ultimately death can occur. AD is the most common form of dementia, accounting for approximately half of all cases. Early-onset familial Alzheimer's disease is a rare form of AD associated with the amyloid precursor protein and presenilin genes. Another form of AD is sporadic AD, which affects over 15 million people worldwide, with its cause primarily unknown. Risk factors for AD include decreased brain size, low education level, low mental ability, head injury, and vascular-disease-related factors \cite{Blennow_2006}. The amyloid hypothesis proposes that extracellular amyloid beta deposits cause AD \cite{hardy1991amyloid}. The tau hypothesis suggests that AD results from tau protein dysfunction, with neurofibrillary tangles formed by tau protein destroying the neuron's transport system \cite{iqbal2005tau}.

Functional magnetic resonance imaging (fMRI) offers a non-invasive approach for diagnosing, evaluating therapeutic interventions, and investigating the mechanisms of AD. In brain imaging, a typical fMRI utilizes the Blood Oxygenation Level Dependent (BOLD) contrast to indirectly reflect brain activity through signal fluctuations. Early studies of brain function primarily relied on task-based fMRI, where fMRI brain activity was acquired during specific functional tasks \cite{pariente2005alzheimer,celone2006alterations}. In 1995, Biswal demonstrated that resting-state fMRI signals could depict spontaneous neuronal activity without the need for external task experiments \cite{biswal1995functional}. An increasing number of studies have utilized resting-state fMRI to investigate brain function and disease-related abnormalities. In recent years, resting-state fMRI has become the most widely used neuroimaging technique in AD-related studies \cite{chen2011classification,wang2006changes,agosta2012resting,binnewijzend2012resting,koch2012diagnostic}.

The analysis of connectivity is a prevalent method in studies related to brain function. It characterizes the interactions between different brain regions in a graph, where the strength of connection quantifies the correlations among them. Structural connectivity can be obtained by applying connectivity analysis to diffusion MRI, where edge weights of the graph are defined as the fibre strength or number. Conversely, functional connectivity analysis can be performed by constructing the functional brain network based on functional MRI. In a typical functional connectivity analysis, the strength of interactions among brain regions is quantified by linear correlations of time series. While structural connectivity depicts the brain's anatomical organization, functional connectivity reveals the co-activation pattern of functionally connected regions, which can be topographically dispersed. Furthermore, functional connectivity provides an approach to assess the dynamic picture of brain activity, unlike structural connectivity, which provides only stationary information on the anatomical profile. It has been reported that only a small portion of functional connections can be explained by underlying structural connections \cite{honey2009predicting,wang2015understanding,mivsic2016network}. Connectivity methods can identify network hubs \cite{achard2006resilient,hagmann2008mapping,buckner2009cortical,sporns2007identification,power2013evidence}, which play a central role in the whole brain network, investigate the modular and hierarchical structure of brain networks \cite{thomas2011organization,power2011functional,buckner2011organization}, and discover disease-related abnormalities in the structure of brain networks \cite{wu2016distinct,zheng2017altered}.

\section{Functional Connectivity Analyses}
The functional connectivity analyses utilized in this study involve two distinct methods: (1) ROI-to-ROI connectivity analysis and (2) ROI-to-Voxel connectivity analysis.

For ROI-to-ROI connectivity analysis, the MRI processing is the same as that for the graph-based analysis, which yields a "de-noised" BOLD signal for each brain region. However, unlike the graph-based method, the threshold is not used to convert the weighted network into a binary network because the weight (strength) of connections is crucial in connectivity analysis. For each pair of ROIs, the connection strength is calculated, and group differences in the strength of connections are analyzed.

Compared to ROI-to-ROI connectivity analysis, the ROI-to-Voxel method is more sophisticated and provides higher analytical resolution. The minimum unit of analysis for the ROI-to-ROI method is the mean BOLD signal of a region, while for the ROI-to-Voxel method, it is the BOLD signal of a voxel. Specifically, a particular region is first selected as the "seed," and its mean BOLD signal is calculated. The functional connections between that seed and all voxels are then quantified by the correlation coefficients of their time series. As a result, a map illustrating the strength of functional connections with the seed can be observed in detail.

\subsection{Data}
\label{section:graphtheory.subject.data}

\subsubsection{Dataset \& Inclusion Criteria}
The participants in our study were obtained from the OASIS-3 public dataset \cite{lamontagne2018oasis}, which is the most recent release of the Open Access Series of Imaging Studies (OASIS). OASIS-3 is a large longitudinal dataset that provides the scientific community with open access not only to multi-modal neuroimaging data but also to various clinical data and cognitive assessments. All data in OASIS-3 are available on the OASIS Brains project website (\url{www.oasis-brains.org}). We employed the same dataset and MR image labelling strategy as a previous study \cite{yao2023altered}. Specifically, clinical diagnoses were used to categorize the MR images into the AD and NC groups. The inclusion criteria are as follows: (1) only data from a single session were downloaded for each individual; (2) the acquisition protocols of BOLD-fMRIs must be the same; (3) for each individual, one BOLD-fMRI must be matched with one T1w MRI from the same session; (4)there must be no significant difference in age, gender, and handedness between the normal control and Alzheimer's Disease group. Supplementary data that can aid in validating this study can be found at \hyperlink{https://github.com/YongchengYAO/AD-FunctionalConnectivity}{https://github.com/YongchengYAO/AD-FunctionalConnectivity}.

\subsubsection{MR Image Acquisition Parameters}
Resting-state BOLD MR images were acquired using a single-shot FID EPI sequence on a 3-Tesla scanner (Siemens, TrioTim or Biograph\textunderscore mMR), with the following parameters: TR = 2200 $ms$; TE = 27 $ms$; FA = $90^{\circ}$; slice thickness = 4 $mm$; slice gap = 0 $mm$; number of slices (z) = 36; in-plane resolution = 4 x 4 $mm^{2}$; in-plane matrix size (x, y) = 64 x 64; number of time points = 164.

T1-weighted MR images were acquired using a single-shot TurboFLASH sequence on the same 3-Tesla scanner (Siemens, TrioTim or Biograph\textunderscore mMR), with the following parameters: TR = 2400 $ms$; TE = 3 $ms$; FA = $8^{\circ}$; slice thickness = 1 $mm$; slice gap = 0 $mm$; number of slices (z) = 176; in-plane resolution = 1 x 1 $mm^{2}$; in-plane matrix size (y, z) = 256 x 256.

\subsubsection{Demographic Information}
The present study involves a total of 590 participants, comprising 175 individuals with AD dementia and 415 normal controls. There is no significant difference in age ($t=1.5125$, $p > 0.05$), gender ($\chi^{2}=2.1782$, $p > 0.05$), and handedness ($\chi^{2}=0.3926$, $p > 0.05$) between the two groups.

\subsection{MR Images Processing}
\begin{figure}[]
	\centering
	\includegraphics[width=0.8\linewidth] {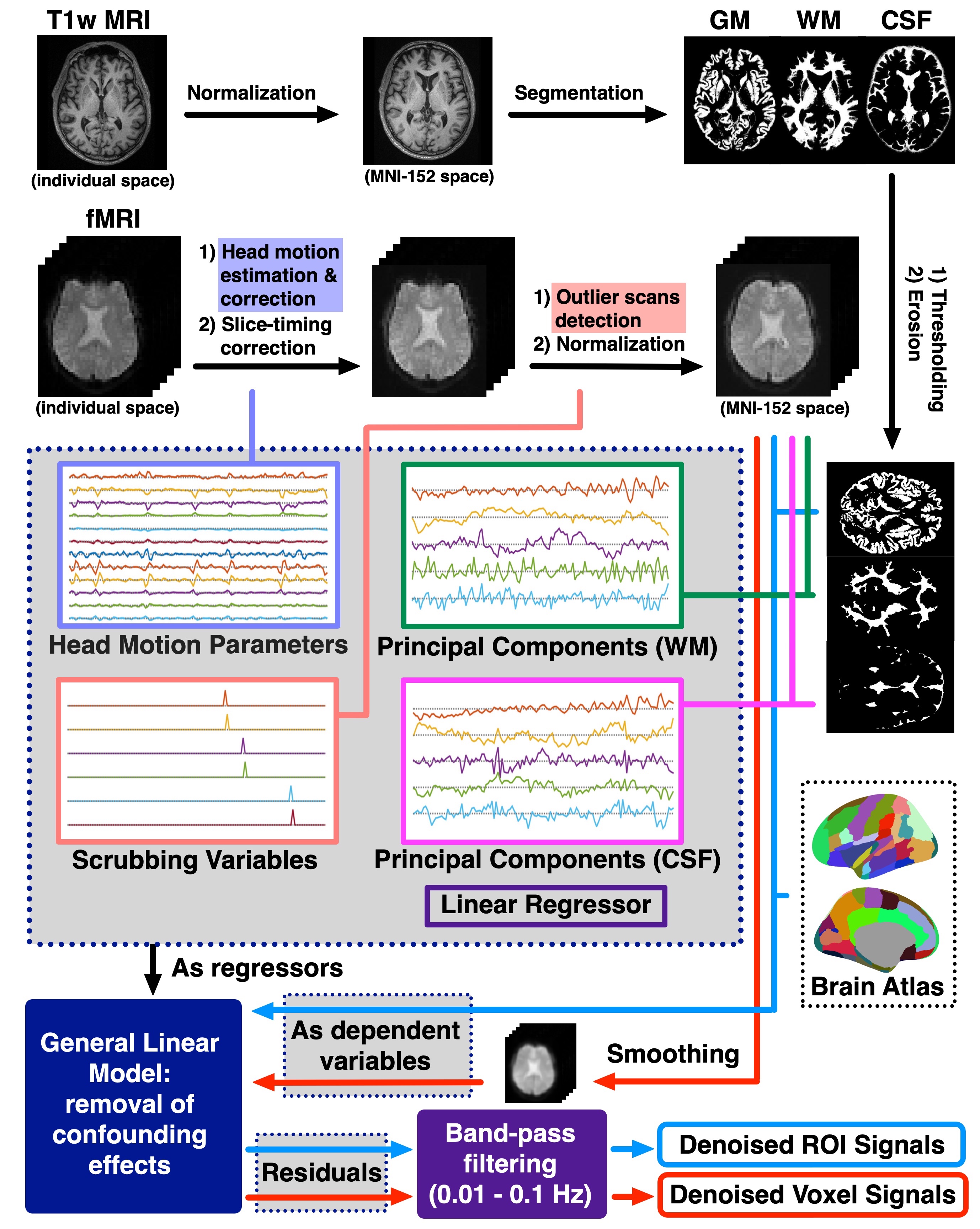}
	\caption[Processing Pipeline of Functional Connectivity Analysis] {{\bf Processing Pipeline of Functional Connectivity Analysis.} The T1-weighted MR images are normalized to the MNI-152 space and then segmented into GM, WM, and CSF. For BOLD fMRI, realignment (head motion estimation and correction), slice timing correction, outlier scan detection, and normalization are employed. Various regressors are used to eliminate confounding effects: (1) Principal components (PCs) from WM and CSF, (2) head motion parameters, (3) scrubbing variables, and (4) a linear regressor. Band-pass filtering ($0.01-0.1$) is applied. Note that the unsmoothed fMRI is used to extract ROI signals before linear regression.}
	\label{fig:pipeline.roi.voxel}
\end{figure}

MRI data were processed using the functional connectivity toolbox (CONN 18.b) \cite{whitfield2012conn}. Figure \ref{fig:pipeline.roi.voxel} illustrates the entire image processing pipeline. 

\paragraph{Normalization and segmentation for T1w MRI.} The T1-weighted MRI was normalized into the MNI-152 space and segmented into grey matter (GM), white matter (WM), and cerebrospinal fluid (CSF). The binary segmentation masks were used to extract the BOLD signal from the normalized (wrapped) functional MRI.

\paragraph{Head motion correction for fMRI.} For the functional MRI, head motion estimation and correction were initially applied to eliminate co-variation across voxels. This is because minor head movements can cause signal disruptions and spurious variance that may either increase or decrease the observed functional connections \cite{power2015recent}. In this study, 6 motion parameters were estimated from the rigid body registration. We included these 6 parameters and their first-order derivatives in a linear regression model to regress out the head-motion-related variance. The head motion correction is also termed "realignment" in literature. Specifically, we registered all other MR volumes in time series to the first volume using B-spline interpolation.

\paragraph{Slice-timing correction for fMRI.} During the acquisition of a BOLD MR image with an EPI sequence, a 3-D volume is effectively a stack of 2-D slices collected one at a time. Therefore, for an fMRI volume at a particular time point, the voxel activations of each slice are not at the same time point. However, it is ideal to observe the activation of the whole brain simultaneously. To address this issue, slice timing correction can be utilized to interpolate slices to a reference slice. This has been demonstrated to be an effective solution that can reliably increase sensitivity and effect power. The implementation details are as follows: (1) the acquisition time for each slice is extracted from the BIDS sidecar that accompanies each NIfTI file; (2) all slices in a volume are interpolated to the slice acquired in the middle of the acquisition time.

\paragraph{Outlier scans detection and normalization for fMRI.} To reduce the effect introduced by severe head motion, we detected outlier scans and created scrubbing variables which were used as regressors in a general linear model. The Artifact Detection Tools (ART) were used for outlier scans detection. The fMRI was registered into the MNI-152 space via non-linear deformation.

\paragraph{Artefacts Removal for fMRI.} Artefacts removal aims to mitigate or eliminate the confounding effects of non-neuronal oscillations due to head movement, cardiac pulsation, respiratory motion, and other systematic noises. Without this step, it is challenging for researchers to determine whether the findings are genuine or driven by artefacts. A general linear model (GLM) was utilized for artefacts removal, with mean BOLD signals extracted from ROIs, and a variety of variables defined as regressors. The nuisance head motion confounding effect can be reduced by regressing out 12 head motion parameters and scrubbing variables. To remove other nuisance effects, the aCompCor method \cite{behzadi2007component} is employed, which is a component-based method with anatomical noise ROIs. Specifically, WM and CSF masks are used to define the WM and CSF areas as noise ROIs. Then, five principal components (PCs) for each noise ROI are calculated via principal component analysis. Lastly, the five PCs from WM and five PCs from CSF are entered into the linear model as regressors. Additionally, the linear trend is removed by adding a linear regressor into the GLM. Finally, the residual time series are band-pass filtered at [$0.01 - 0.1$] Hz to retain neuroactivity-related intrinsic signal fluctuations.

\paragraph{BOLD signal.} Two types of BOLD signals are extracted: ROI signals, which are the mean BOLD signals of all voxels within a pre-defined ROI, and voxel signals, which are the de-noised BOLD signals of each voxel. The unsmoothed functional MR image is used to extract ROI signals (mean signal) that are further de-noised using a general linear model and band-pass filtering (0.01-0.1 Hz). To de-noise voxel signals, spatial smoothing is first applied to the functional MR image. For each voxel signal in the smoothed image, linear regression and band-pass filtering are used to remove confounding effects.

% Figure: Pre-processing Pipeline.
%——————————————————————
%——————————————————————

\subsection{ROI-to-ROI Connectivity Analysis}
\subsubsection{Definition of Functional Connections}
\begin{figure}[]
	\centering
	\includegraphics[width=0.9\linewidth] {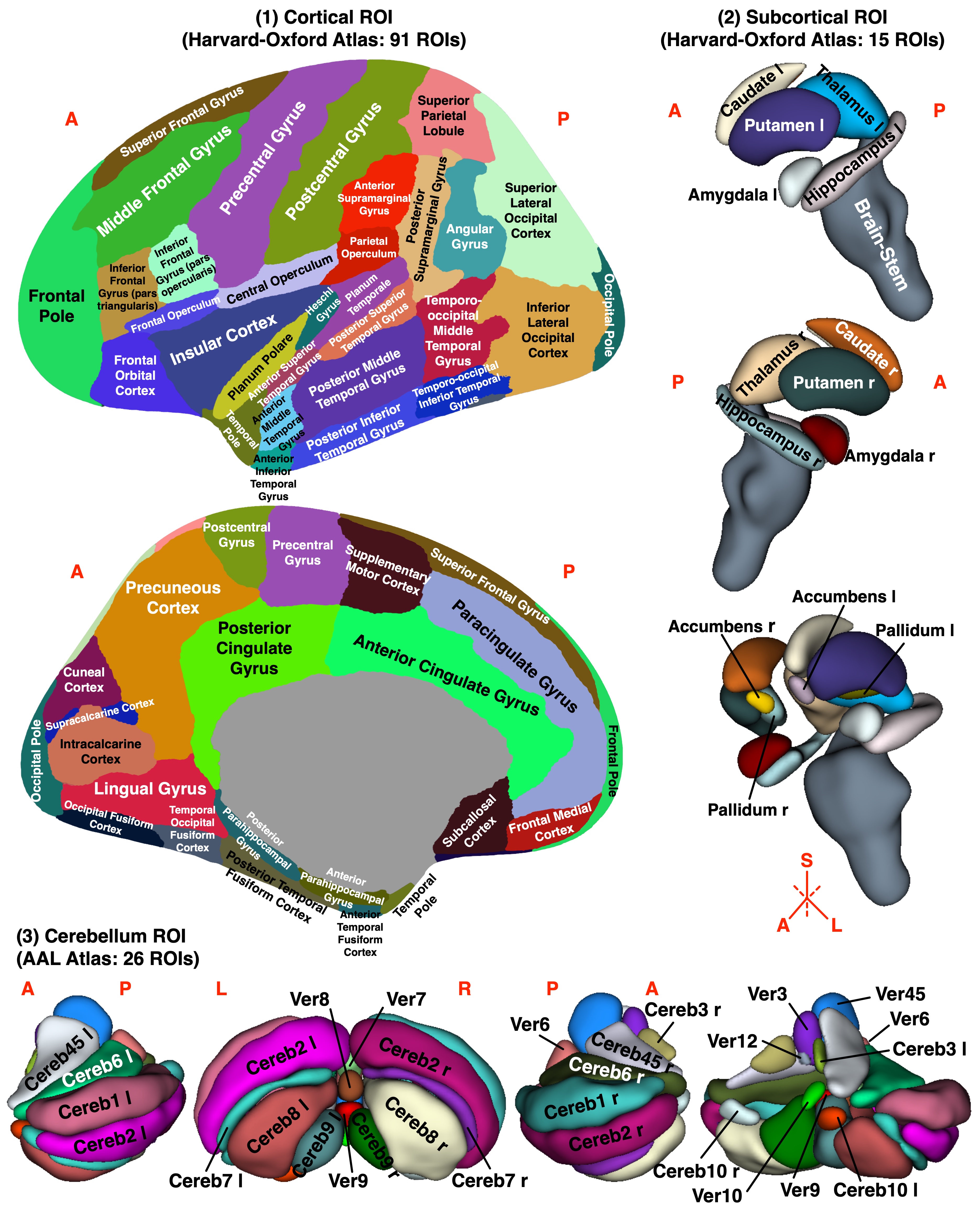}
	\caption[Brain Atlas] {{\bf Brain Atlas.} It is a customized brain regions parcellation scheme, combining (1) 91 cortical regions (from the Harvard-Oxford Atlas), (2) 15 subcortical structures (from the Harvard-Oxford Atlas), and (3) 26 cerebellum regions (from the AAL Atlas). In the 3-D rendering of subcortical and cerebellum regions, spatial smoothing is applied for better visualization. (A: anterior; P: posterior; L: left; R: right; S: superior)}
	\label{fig:brain.atlas}
\end{figure}

The processing pipeline produces "de-noised" ROI signals, which are mean signals for each ROI defined in the "Harvard-Oxford-AAL" atlas (Figure \ref{fig:brain.atlas}). The functional connection between a pair of ROIs is defined as the Pearson's linear correlation coefficient, and the Fisher z-transformation is applied to the correlation coefficient.

\subsubsection{Statistical Analysis}
Initially, the functional connections (FCs) for all subjects are computed. Subsequently, two-sample two-tailed t-tests are employed to compare each FC between the AD and NC groups. Finally, false discovery rate (FDR) correction is applied to correct for multiple comparisons across all FCs (number of FCs: $132*131/2=8646$).

% Figure: Brain Atlas. 
%——————————————————————
%——————————————————————

\subsection{ROI-to-Voxel Connectivity Analysis}
\subsubsection{Seed Region \& Seed-to-Voxel FC Map}
ROI-to-Voxel connectivity is also known as seed-to-voxel connectivity, which generates voxel-level functional connectivity maps for each seed region. The first step is to define the seed region, which can be any ROI defined by researchers, such as a manually drawn ROI or an ROI of a small sphere at a specific position. In this study, the seed ROI is defined by the "Harvard-Oxford-AAL" atlas (Figure \ref{fig:brain.atlas}). For each seed region, the voxel-level functional connectivity map is generated by calculating the FCs between the seed and all other voxels. The seed-to-voxel FC is also defined as the Fisher z-transformed Pearson's linear correlation coefficient.

\subsubsection{Statistical Analysis}
Initially, all seed-to-voxel FCs (number = the number of voxels) for all seed regions (number = 132) and subjects (number = 590) are computed. Subsequently, two-sample t-tests are used to compare each voxel-level FC between the AD and NC groups. Thirdly, FCs surviving both connection-level threshold (uncorrected $p <0.001$) and cluster-size threshold (FDR-corrected $p <0.05$) are considered statistically significant. The cluster-size threshold is motivated by the idea that spuriously significant voxels are unlikely to form a large cluster. Therefore, techniques like random field theory can be used to form a null model for cluster size, which provides the probability $P_n$ of observing a cluster of size larger or equal to $n$ under random activation. In the seed-to-voxel FC analysis, the chosen FDR-corrected p-value is related to a cluster size in the null model, and that cluster size is used as a threshold.

\subsection{Results}
\subsubsection{ROI-to-ROI Functional Connectivity}
\begin{figure}[]
	\centering
	\includegraphics[width=0.9\linewidth]{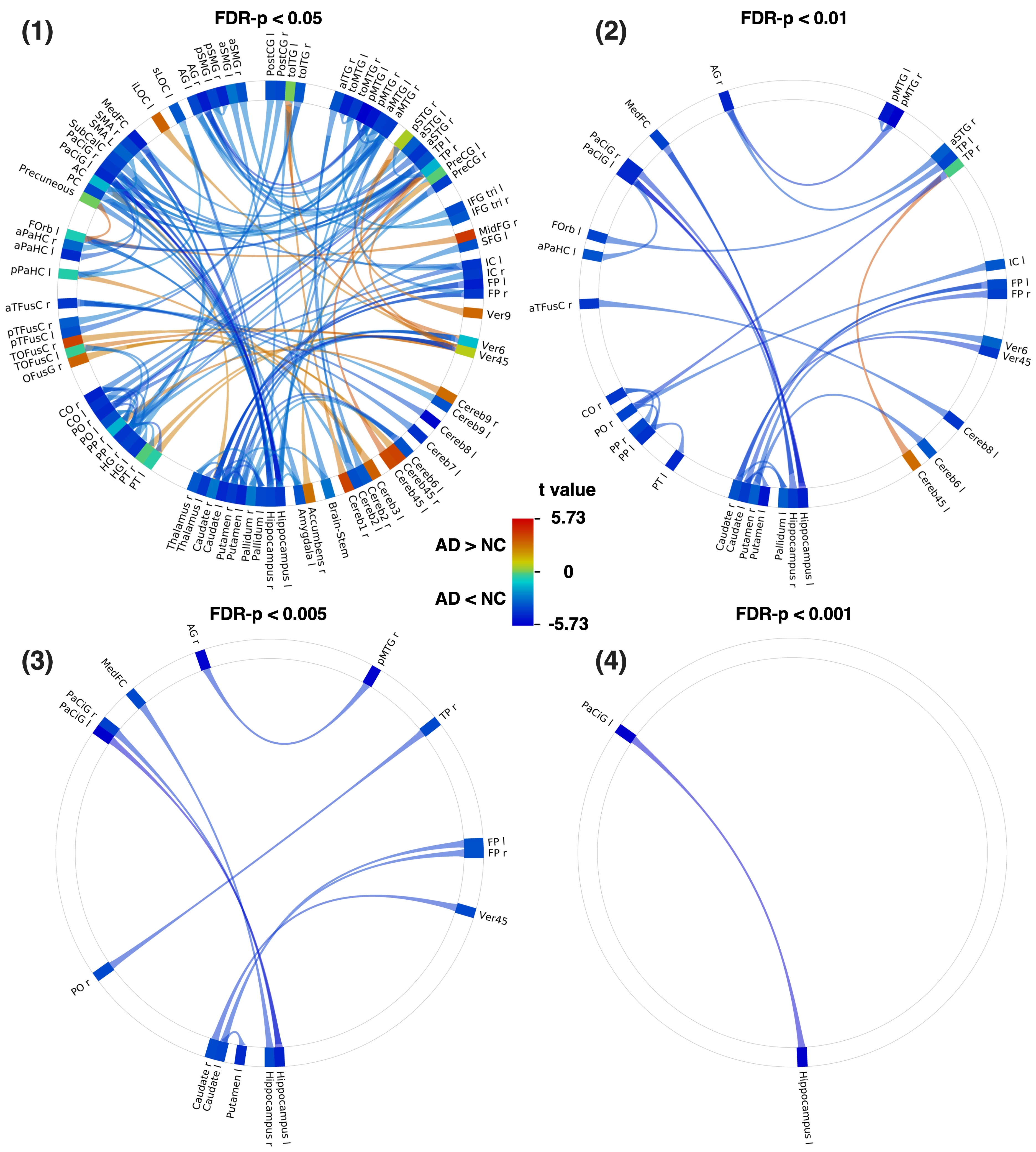}
	\caption[Altered ROI-to-ROI Functional Connectivity in AD] {{\bf Altered ROI-to-ROI Functional Connectivity in AD.} Widespread decreased functional connections are found at different significant level: (1) FDR-p $< 0.05$, (2) FDR-p $< 0.01$, (3) FDR-p $< 0.005$, (4) FDR-p $< 0.001$}
	\label{fig:roi2roi.ring}
\end{figure}

Table \ref{tab:result.roi2roi} displays the functional connections that are significantly different in the AD group compared to the NC group. Widespread decreased functional connections (Table \ref{tab:result.roi2roi} and Figure \ref{fig:roi2roi.ring}) are observed in the AD group. FCs that survive various thresholds (FDR-corrected p-value $<0.05$, $0.01$, $0.005$, and $0.001$) are explored and compared (Figure \ref{fig:roi2roi.ring}). The decreased FCs in AD have a larger effect size than the increased FCs. Below is a summary of the findings.

%list
%———————————————————————
\begin{itemize}
\item Under a loose threshold (FDR-p $<0.05$) (Figure \ref{fig:roi2roi.ring} (1)), both decreased and increased FCs are observed in the AD group compared to the NC group.
\item Under a moderate threshold (FDR-p $<0.01$) (Figure \ref{fig:roi2roi.ring} (2)), a variety of decreased FCs are observed. However, only one increased FC (between the right TP and left Cereb45) is observed.
\item Under a relatively stringent threshold (FDR-p $<0.005$), only decreased FCs are found in the AD group: left PaCiG to left Hippocampus (FDR-p $=0.0001$), Ver45 to right Caudate (FDR-p $=0.0030$), right pMTG to right AG (FDR-p $=0.0030$), right TP to right PO (FDR-p $=0.0030$), right PaCiG to left Hippocampus (FDR-p $=0.0030$), MedFC to right Hippocampus (FDR-p $=0.0037$), right FP to left Caudate (FDR-p $=0.0037$), left FP to left Caudate (FDR-p $=0.0037$), and left Putamen to left Caudate (FDR-p $=0.0049$).
\item Under the most stringent threshold (FDR-p $<0.001$), only the decreased functional connection between the left PaCiG to left Hippocampus (FDR-p $=0.0001$) is statistically significant in the AD group.
\end{itemize}
%———————————————————————

%Tab. Title
%——————————————————————
\begin{center}
\begin{small}
\setlength\LTleft{0cm}
\setlength\LTright{0cm}
\begin{longtable}[c]{l | ccc}
	\caption{Altered Functional Connections in AD} \label{tab:result.roi2roi} \\
% endfirsthead
\hline \hline 
\multicolumn{1}{c}{\textbf{Functional Connection}} & \multicolumn{1}{c}{\textbf{Statistic}} & \multicolumn{1}{c}{\textbf{p}} & \multicolumn{1}{c}{\textbf{FDR-p}}  \\ 
\hline \hline 
\endfirsthead
%endhead
\multicolumn{4}{l}%
{{\tablename\ \thetable{} -- continued from previous page}} \\
\hline 
\multicolumn{1}{c}{\textbf{Functional Connection}} & \multicolumn{1}{c}{\textbf{Statistic}} & \multicolumn{1}{c}{\textbf{p}} & \multicolumn{1}{c}{\textbf{FDR-p}}  \\
\hline 
\endhead
%endfoot
\hline \multicolumn{4}{r}{{Continued on next page}} \\ \hline
\endfoot
%endlastfoot
\hline \hline
\endlastfoot
% the content of long table
% Table generated by Excel2LaTeX from sheet 'Sheet3'
PaCiG l - Hippocampus l & T(588) = -5.73 & $<$ 0.0001 & 0.0001 \bigstrut[t]\\
Ver45 - Caudate r & T(588) = -4.86 & $<$ 0.0001 & 0.0030 \\
pMTG r - AG r & T(588) = -4.85 & $<$ 0.0001 & 0.0030 \\
TP r - PO r & T(588) = -4.83 & $<$ 0.0001 & 0.0030 \\
PaCiG r - Hippocampus l & T(588) = -4.83 & $<$ 0.0001 & 0.0030 \\
MedFC - Hippocampus r & T(588) = -4.74 & $<$ 0.0001 & 0.0037 \\
FP r - Caudate l & T(588) = -4.69 & $<$ 0.0001 & 0.0037 \\
FP l - Caudate l & T(588) = -4.69 & $<$ 0.0001 & 0.0037 \\
Putamen l - Caudate l & T(588) = -4.60 & $<$ 0.0001 & 0.0049 \\
PaCiG l - Hippocampus r & T(588) = -4.50 & $<$ 0.0001 & 0.0069 \\
PT l - PP l & T(588) = -4.48 & $<$ 0.0001 & 0.0069 \\
PP l - PO r & T(588) = -4.44 & $<$ 0.0001 & 0.0069 \\
Cereb8 l - aTFusC r & T(588) = -4.44 & $<$ 0.0001 & 0.0069 \\
Pallidum l - Caudate l & T(588) = -4.43 & $<$ 0.0001 & 0.0069 \\
Putamen r - Cereb6 l & T(588) = -4.42 & $<$ 0.0001 & 0.0069 \\
TP l - FOrb l & T(588) = -4.40 & $<$ 0.0001 & 0.0069 \\
MedFC - Hippocampus l & T(588) = -4.38 & $<$ 0.0001 & 0.0069 \\
PP l - CO r & T(588) = -4.37 & $<$ 0.0001 & 0.0069 \\
PO r - IC l & T(588) = -4.29 & $<$ 0.0001 & 0.0083 \\
pMTG l - pMTG r & T(588) = -4.29 & $<$ 0.0001 & 0.0083 \\
PaCiG l - aPaHC l & T(588) = -4.29 & $<$ 0.0001 & 0.0083 \\
aSTG r - AG r & T(588) = -4.29 & $<$ 0.0001 & 0.0083 \\
Ver6 - Putamen r & T(588) = -4.28 & $<$ 0.0001 & 0.0083 \\
Pallidum l - Caudate r & T(588) = -4.26 & $<$ 0.0001 & 0.0085 \\
PP r - CO r & T(588) = -4.22 & $<$ 0.0001 & 0.0092 \\
FP l - Caudate r & T(588) = -4.22 & $<$ 0.0001 & 0.0092 \\
Ver45 - Caudate l & T(588) = -4.18 & $<$ 0.0001 & 0.0102 \\
Thalamus l - Caudate l & T(588) = -4.17 & $<$ 0.0001 & 0.0103 \\
aMTG r - AG r & T(588) = -4.15 & $<$ 0.0001 & 0.0110 \\
toITG r - pTFusC l & T(588) = -4.13 & $<$ 0.0001 & 0.0115 \\
Putamen l - Caudate r & T(588) = -4.08 & 0.0001 & 0.0128 \\
PP l - PreCG r & T(588) = -4.04 & 0.0001 & 0.0147 \\
Hippocampus r - sLOC l & T(588) = -4.04 & 0.0001 & 0.0147 \\
aTFusC r - Cereb7 l & T(588) = -4.03 & 0.0001 & 0.0149 \\
PaCiG r - aMTG r & T(588) = -4.02 & 0.0001 & 0.0149 \\
pMTG l - aMTG r & T(588) = -3.98 & 0.0001 & 0.0167 \\
Hippocampus l - FOrb l & T(588) = -3.97 & 0.0001 & 0.0167 \\
PC - Hippocampus l & T(588) = -3.96 & 0.0001 & 0.0171 \\
CO r - TP r & T(588) = -3.94 & 0.0001 & 0.0178 \\
AC - IC l & T(588) = -3.94 & 0.0001 & 0.0176 \\
Putamen l - Ver6 & T(588) = -3.91 & 0.0001 & 0.0181 \\
Hippocampus l - sLOC l & T(588) = -3.91 & 0.0001 & 0.0181 \\
FP r - aMTG r & T(588) = -3.91 & 0.0001 & 0.0181 \\
pPaHC l - PaCiG l & T(588) = -3.89 & 0.0001 & 0.0191 \\
TP r - CO l & T(588) = -3.88 & 0.0001 & 0.0193 \\
Cereb2 l - aITG r & T(588) = -3.88 & 0.0001 & 0.0193 \\
Putamen l - Cereb6 l & T(588) = -3.87 & 0.0001 & 0.0194 \\
Precuneous - Hippocampus r & T(588) = -3.83 & 0.0001 & 0.0224 \\
Hippocampus l - aMTG r & T(588) = -3.83 & 0.0001 & 0.0224 \\
SubCalC - Hippocampus l & T(588) = -3.82 & 0.0001 & 0.0226 \\
toMTG r - toMTG l & T(588) = -3.81 & 0.0002 & 0.0226 \\
HG r - CO r & T(588) = -3.81 & 0.0002 & 0.0226 \\
Thalamus r - Caudate l & T(588) = -3.79 & 0.0002 & 0.0237 \\
SMA r - PP l & T(588) = -3.78 & 0.0002 & 0.0242 \\
CO l - PP l & T(588) = -3.77 & 0.0002 & 0.0242 \\
AG l - pMTG r & T(588) = -3.77 & 0.0002 & 0.0242 \\
aITG r - Cereb2 r & T(588) = -3.76 & 0.0002 & 0.0243 \\
PP l - TOFusC l & T(588) = -3.75 & 0.0002 & 0.0245 \\
Caudate l - Putamen r & T(588) = -3.75 & 0.0002 & 0.0245 \\
SubCalC - TP l & T(588) = -3.74 & 0.0002 & 0.0245 \\
pSTG r - TP r & T(588) = -3.74 & 0.0002 & 0.0245 \\
Caudate r - FP r & T(588) = -3.74 & 0.0002 & 0.0245 \\
AC - Hippocampus l & T(588) = -3.74 & 0.0002 & 0.0245 \\
Hippocampus l - SFG l & T(588) = -3.72 & 0.0002 & 0.0255 \\
FOrb l - TP r & T(588) = -3.72 & 0.0002 & 0.0255 \\
Hippocampus r - PC & T(588) = -3.71 & 0.0002 & 0.0259 \\
aMTG l - PC & T(588) = -3.71 & 0.0002 & 0.0259 \\
Amygdala l - PaCiG l & T(588) = -3.69 & 0.0002 & 0.0270 \\
aMTG l - pMTG r & T(588) = -3.69 & 0.0002 & 0.0273 \\
aSMG l - Brain-Stem & T(588) = -3.68 & 0.0002 & 0.0273 \\
pSMG l - pSMG r & T(588) = -3.68 & 0.0003 & 0.0273 \\
Hippocampus r - PaCiG r & T(588) = -3.68 & 0.0003 & 0.0273 \\
AG l - TP l & T(588) = -3.66 & 0.0003 & 0.0289 \\
Hippocampus r - SubCalC & T(588) = -3.62 & 0.0003 & 0.0321 \\
PO r - TP l & T(588) = -3.61 & 0.0003 & 0.0331 \\
PO l - PP r & T(588) = -3.60 & 0.0003 & 0.0339 \\
aMTG r - MedFC & T(588) = -3.60 & 0.0003 & 0.0341 \\
PostCG r - PP l & T(588) = -3.59 & 0.0004 & 0.0343 \\
Brain-Stem - Pallidum r & T(588) = -3.58 & 0.0004 & 0.0351 \\
toITG r - pTFusC r & T(588) = -3.57 & 0.0004 & 0.0362 \\
TOFusC l - PP r & T(588) = -3.57 & 0.0004 & 0.0362 \\
Thalamus r - Pallidum l & T(588) = -3.56 & 0.0004 & 0.0362 \\
PP l - IC r & T(588) = -3.56 & 0.0004 & 0.0362 \\
pMTG r - aMTG r & T(588) = -3.56 & 0.0004 & 0.0362 \\
Cereb2 l - aTFusC r & T(588) = -3.56 & 0.0004 & 0.0362 \\
Ver45 - Putamen r & T(588) = -3.55 & 0.0004 & 0.0363 \\
PaCiG l -TP l & T(588) = -3.55 & 0.0004 & 0.0365 \\
Ver6 - Caudate r & T(588) = -3.54 & 0.0004 & 0.0369 \\
toITG l - Hippocampus r & T(588) = -3.54 & 0.0004 & 0.0370 \\
aSMG r - IC l & T(588) = -3.54 & 0.0004 & 0.0369 \\
HG l - PP l & T(588) = -3.53 & 0.0004 & 0.0373 \\
aPaHC r - PaCiG r & T(588) = -3.53 & 0.0004 & 0.0373 \\
TP r - PaCiG r & T(588) = -3.52 & 0.0005 & 0.0377 \\
PP l - PostCG l & T(588) = -3.52 & 0.0005 & 0.0380 \\
PaCiG r - TP r & T(588) = -3.52 & 0.0005 & 0.0377 \\
Hippocampus l - PreCG r & T(588) = -3.52 & 0.0005 & 0.0380 \\
Hippocampus r - FOrb l & T(588) = -3.51 & 0.0005 & 0.0385 \\
Caudate l - Amygdala l & T(588) = -3.51 & 0.0005 & 0.0380 \\
TP l - pSMG l & T(588) = -3.50 & 0.0005 & 0.0385 \\
PP l - PO l & T(588) = -3.50 & 0.0005 & 0.0385 \\
Cereb2 l - aMTG l & T(588) = -3.50 & 0.0005 & 0.0385 \\
PreCG l - PP l & T(588) = -3.49 & 0.0005 & 0.0391 \\
PC - aSTG l & T(588) = -3.48 & 0.0005 & 0.0400 \\
Hippocampus r - Hippocampus l & T(588) = -3.47 & 0.0005 & 0.0406 \\
Cereb9 l - aTFusC r & T(588) = -3.46 & 0.0006 & 0.0428 \\
SMA r - HG r & T(588) = -3.44 & 0.0006 & 0.0446 \\
TP l - AG r & T(588) = -3.43 & 0.0006 & 0.0448 \\
Pallidum r - Cereb6 l & T(588) = -3.43 & 0.0006 & 0.0448 \\
PaCiG l - aMTG r & T(588) = -3.43 & 0.0006 & 0.0448 \\
aMTG r - PaCiG l & T(588) = -3.43 & 0.0006 & 0.0448 \\
toMTG r - IFG tri r & T(588) = -3.43 & 0.0007 & 0.0448 \\
IFG tri r - toMTG r & T(588) = -3.43 & 0.0007 & 0.0448 \\
aMTG r - AG l & T(588) = -3.43 & 0.0007 & 0.0448 \\
Putamen r - Putamen l & T(588) = -3.42 & 0.0007 & 0.0448 \\
PT r - PP l & T(588) = -3.42 & 0.0007 & 0.0448 \\
PC - aMTG r & T(588) = -3.42 & 0.0007 & 0.0448 \\
Hippocampus l - Amygdala l & T(588) = -3.41 & 0.0007 & 0.0463 \\
toITG l - IFG tri l & T(588) = -3.40 & 0.0007 & 0.0463 \\
SMA l - AC & T(588) = -3.40 & 0.0007 & 0.0463 \\
sLOC l - pPaHC l & T(588) = -3.40 & 0.0007 & 0.0463 \\
pPaHC l - sLOC l & T(588) = -3.40 & 0.0007 & 0.0463 \\
pMTG r - Hippocampus l & T(588) = -3.40 & 0.0007 & 0.0463 \\
OFusG r - Cereb3 l & T(588) = 3.39 & 0.0007 & 0.0464 \\
Caudate r - AC & T(588) = 3.39 & 0.0007 & 0.0464 \\
Ver45 - PT l & T(588) = 3.40 & 0.0007 & 0.0463 \\
PreCG l - Accumbens r & T(588) = 3.40 & 0.0007 & 0.0463 \\
TOFusC r - Cereb45 r & T(588) = 3.44 & 0.0006 & 0.0444 \\
Ver45 - PT r & T(588) = 3.45 & 0.0006 & 0.0434 \\
OFusG r - Cereb45 r & T(588) = 3.47 & 0.0005 & 0.0406 \\
Ver45 - iLOC l & T(588) = 3.50 & 0.0005 & 0.0385 \\
Ver45 - TOFusC r & T(588) = 3.57 & 0.0004 & 0.0362 \\
pPaHC l - Cereb3 l & T(588) = 3.63 & 0.0003 & 0.0316 \\
toITG l - Cereb1 r & T(588) = 3.68 & 0.0003 & 0.0273 \\
Ver9 - pSTG r & T(588) = 3.77 & 0.0002 & 0.0242 \\
PP r - Cereb9 r & T(588) = 3.79 & 0.0002 & 0.0237 \\
MidFG r - FOrb l & T(588) = 3.81 & 0.0002 & 0.0226 \\
Ver45 - TP r & T(588) = 3.91 & 0.0001 & 0.0181 \\
Ver45 - TOFusC l & T(588) = 3.98 & 0.0001 & 0.0167 \\
Ver45 - pSTG r & T(588) = 3.98 & 0.0001 & 0.0167 \\
Precuneous - FOrb l & T(588) = 4.09 & $<$ 0.0001 & 0.0128 \\
Ver6 - toITG l & T(588) = 4.12 & $<$ 0.0001 & 0.0116 \\
TP r - Cereb45 l & T(588) = 4.22 & $<$ 0.0001 & 0.0092 \bigstrut[b]\\
\end{longtable}
\end{small}
\end{center}
%——————————————————————

% Figure: Altered ROI-to-ROI Functional Connectivity in AD
%——————————————————————
%——————————————————————

\subsubsection{ROI-to-Voxel Functional Connectivity Map}
\begin{figure}[]
	\centering
	\includegraphics[width=1\linewidth] {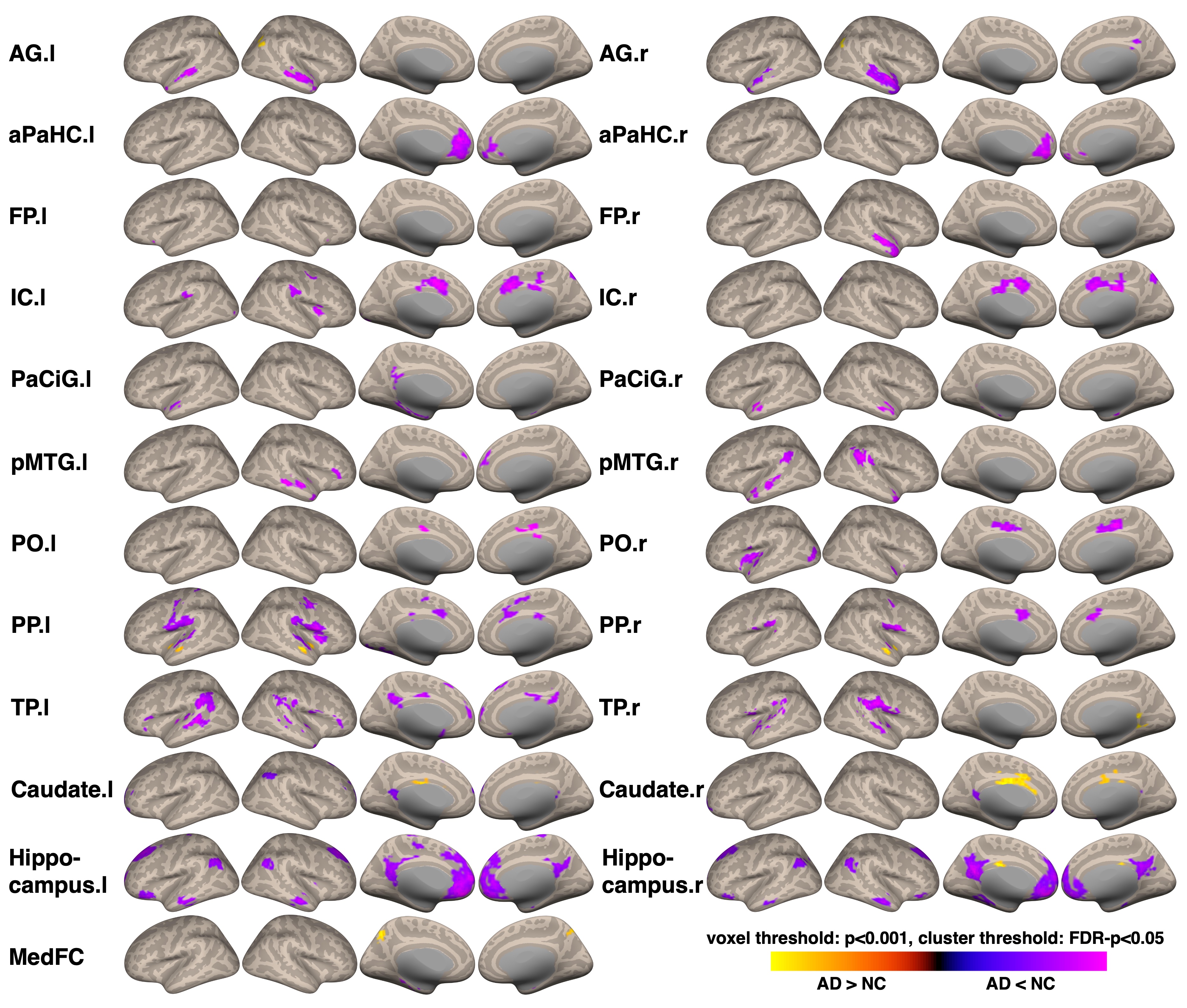}
	\caption[Voxel-level Connectivity Maps of 23 Seed Regions] {{\bf Voxel-level Connectivity Maps of 23 Seed Regions.} l: left; r: right; AG: Angular Gyrus; aPaHC: anterior Parahippocampal Gyrus; FP: Frontal Pole; IC: Insular Cortex; PaCiG: Paracingulate Gyrus; pMTG: posterior Middle Temporal Gyrus; PO: Parietal Operculum Cortex; PP: Planum Polare; TP: Temporal Pole; MedFC: Frontal Medial Cortex; AD: Alzheimer’s Disease; NC: Normal Controls.}
	\label{fig:fcmap.all}
\end{figure}

The ROI-to-voxel (also known as seed-to-voxel) functional connections generate an FC map for a seed region. In this study, the seeds are 132 ROIs defined in the Harvard-Oxford-AAl atlas. Several seed-to-voxel FC maps for ROIs are presented here (Figure \ref{fig:fcmap.all}), including the bilateral Hippocampus (Figure \ref{fig:fcmap.hippocampus}), bilateral anterior Parahippocampal Gyrus (aPaHC, Figure \ref{fig:fcmap.aPaHC}), bilateral Angular Gyrus (AG, Figure \ref{fig:fcmap.AG}), bilateral posterior Middle Temporal Gyrus (pMTG, Figure \ref{fig:fcmap.pMTG}), bilateral Temporal Pole (TP, Figure \ref{fig:fcmap.TP}), bilateral Insular Cortex (IC, Figure \ref{fig:fcmap.IC}), bilateral Planum Polare (PP, Figure \ref{fig:fcmap.PP}), bilateral Paracingulate Gyrus (PaCiG, Figure \ref{fig:fcmap.PaCiG}), bilateral Parietal Operculum (PO, Figure \ref{fig:fcmap.PO}), bilateral Frontal Pole (FP, Figure \ref{fig:fcmap.FP}), Frontal Medial Cortex (MedFC, Figure \ref{fig:fcmap.MedFC}), and bilateral Caudate (Figure \ref{fig:fcmap.Caudate}).

% Figure: all FC map
%——————————————————————
%——————————————————————

\begin{itemize}
\item In Figure \ref{fig:fcmap.hippocampus}, the functional connections that are significantly decreased in the AD group with the left Hippocampus are mainly located in the bilateral Paracingulate Gyrus (PaCiG), bilateral Frontal Pole (FP), Frontal Medial Cortex (MedFC), Subcallosal Cortex (SubCalC), bilateral Superior Frontal Gyrus (SFG), Anterior Cingulate Gyrus (AC), Posterior Cingulate Gyrus (PC), Precuneous, bilateral Precentral Gyrus (PreCG), bilateral Angular Gyrus (AG), bilateral superior Lateral Occipital Cortex (sLOC), bilateral posterior Middle Temporal Gyrus (pMTG), bilateral posterior Inferior Temporal Gyrus (pITG), and left Frontal Orbital Cortex (FOrb). It is important to note that only the results for cortical cortices are visible in the surface mapping of test statistics (3D view in Figure \ref{fig:fcmap.hippocampus}). The axial view (axial slices in dotted square in Figure \ref{fig:fcmap.hippocampus}) provides additional information on the functional connectivity at subcortical structures and cerebellum. It is evident that there is a significant decreased FC between the left Hippocampus (as seed ROI) and the right Hippocampus.

\item Similarly, in Figure \ref{fig:fcmap.hippocampus}, the right Hippocampus exhibits a similar FC map to the left Hippocampus (Figure \ref{fig:fcmap.hippocampus}). The significant clusters are located in the bilateral Paracingulate Gyrus (PaCiG), bilateral Frontal Pole (FP), Frontal Medial Cortex (MedFC), Subcallosal Cortex (SubCalC), bilateral Superior Frontal Gyrus (SFG), Anterior Cingulate Gyrus (AC), Posterior Cingulate Gyrus (PC), Precuneous, bilateral Angular Gyrus (AG), bilateral superior Lateral Occipital Cortex (sLOC), bilateral posterior Middle Temporal Gyrus (pMTG), bilateral Frontal Orbital Cortex (FOrb), left posterior Inferior Temporal Gyrus (pITG), left posterior Parahippocampal Gyrus (pPaHC), left Temporal Occipital Fusiform Cortex (TOFusC), and bilateral posterior Temporal Fusiform Cortex (pTFusC).

% Figure: Voxel-level Connectivity Map of Hippocampus.
%——————————————————————
\begin{figure}[]
	\centering
	\includegraphics[width=1\linewidth]{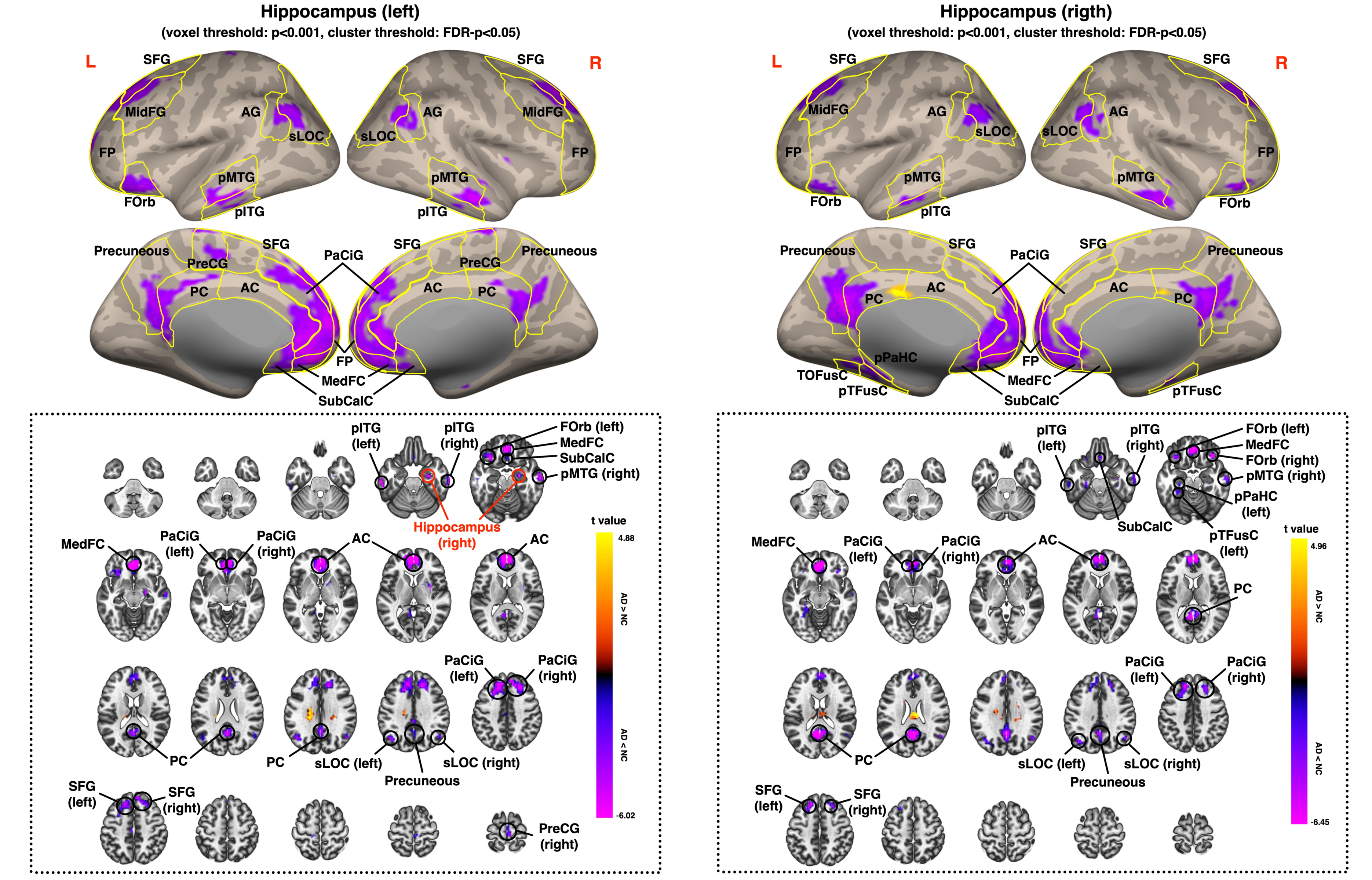}
	\caption[Voxel-level Connectivity Map of Left Hippocampus] {{\bf Voxel-level Connectivity Map of Left Hippocampus.} }
	\label{fig:fcmap.hippocampus}
\end{figure}
%——————————————————————

\item Figure \ref{fig:fcmap.aPaHC} demonstrates that the functional connections that are significantly decreased in the AD group with the left or right anterior Parahippocampal Gyrus (aPaHC) are primarily located in the left Paracingulate Gyrus (PaCiG) and Anterior Cingulate Gyrus (AC).

% Figure: Voxel-level Connectivity Map of aPaHC.
%——————————————————————
\begin{figure}[]
	\centering
	\includegraphics[width=0.9\linewidth]{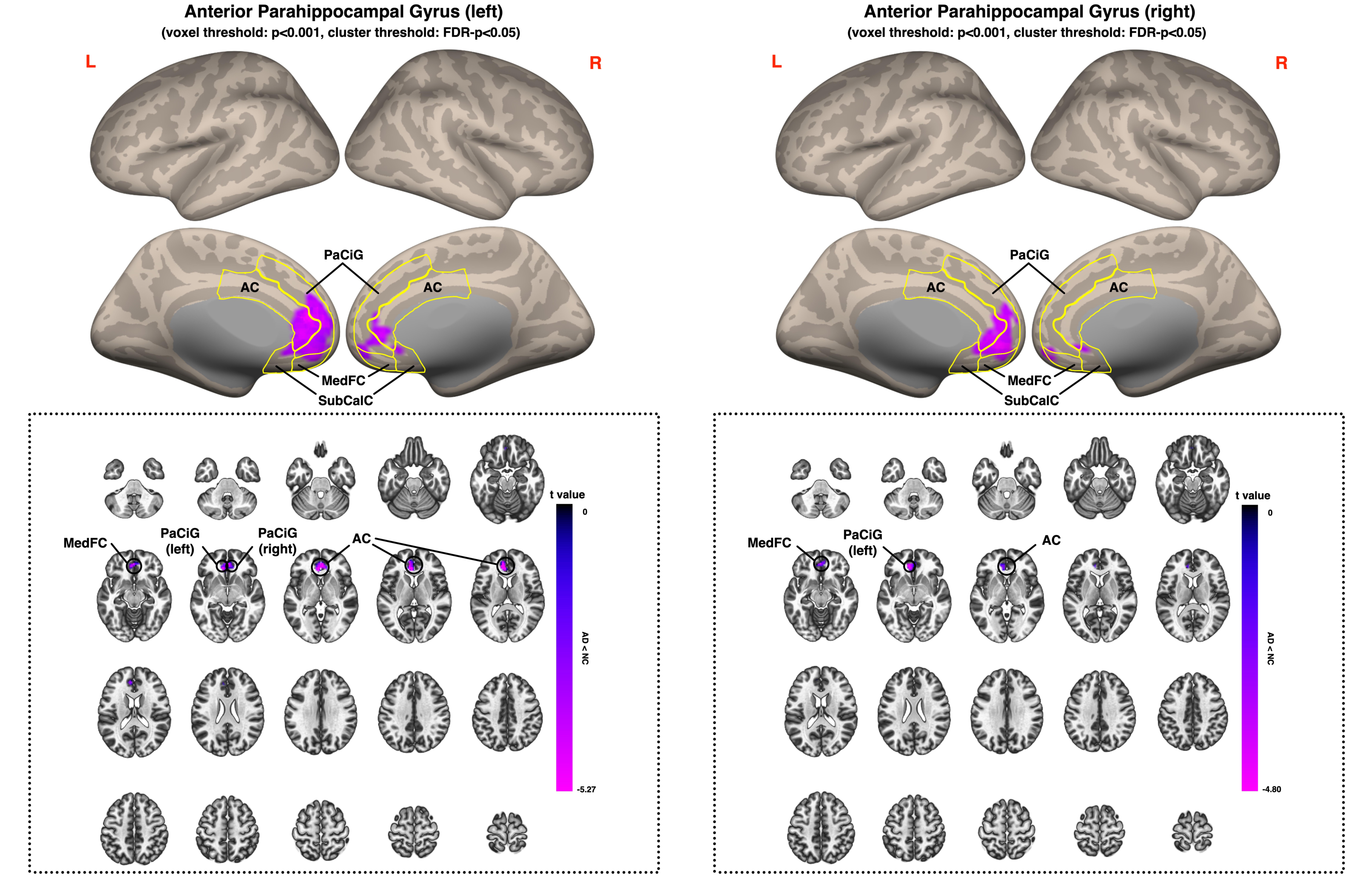}
	\caption[Voxel-level Connectivity Map of Left Anterior Parahippocampal Gyrus] {{\bf Voxel-level Connectivity Map of Left Anterior Parahippocampal Gyrus.} }
	\label{fig:fcmap.aPaHC}
\end{figure}
%——————————————————————

\item are primarily located in bilateral posterior Middle Temporal Gyrus (pMTG), right anterior Middle Temporal Gyrus (aMTG), and bilateral Temporal Pole (TP). Moreover, increased functional connections are observed in the right superior Lateral Occipital Cortex (sLOC).

\item Similarly, in Figure \ref{fig:fcmap.AG}, the functional connections that are significantly decreased in the AD group with the right Angular Gyrus (AG) are primarily located in the right temporal gyrus (including the pMTG, aMTG, aITG, aSTG, pSTG, and TP), right Planum Polare (PP), and Posterior Cingulate Gyrus (PC). Furthermore, decreased FCs are also observed in the left hemisphere (including the pSTG, pMTG, aMTG, and TP). Additionally, increased functional connections are observed in the right superior Lateral Occipital Cortex (sLOC).

% Figure: Voxel-level Connectivity Map of Left AG.
%——————————————————————
\begin{figure}[]
	\centering
	\includegraphics[width=0.9\linewidth]{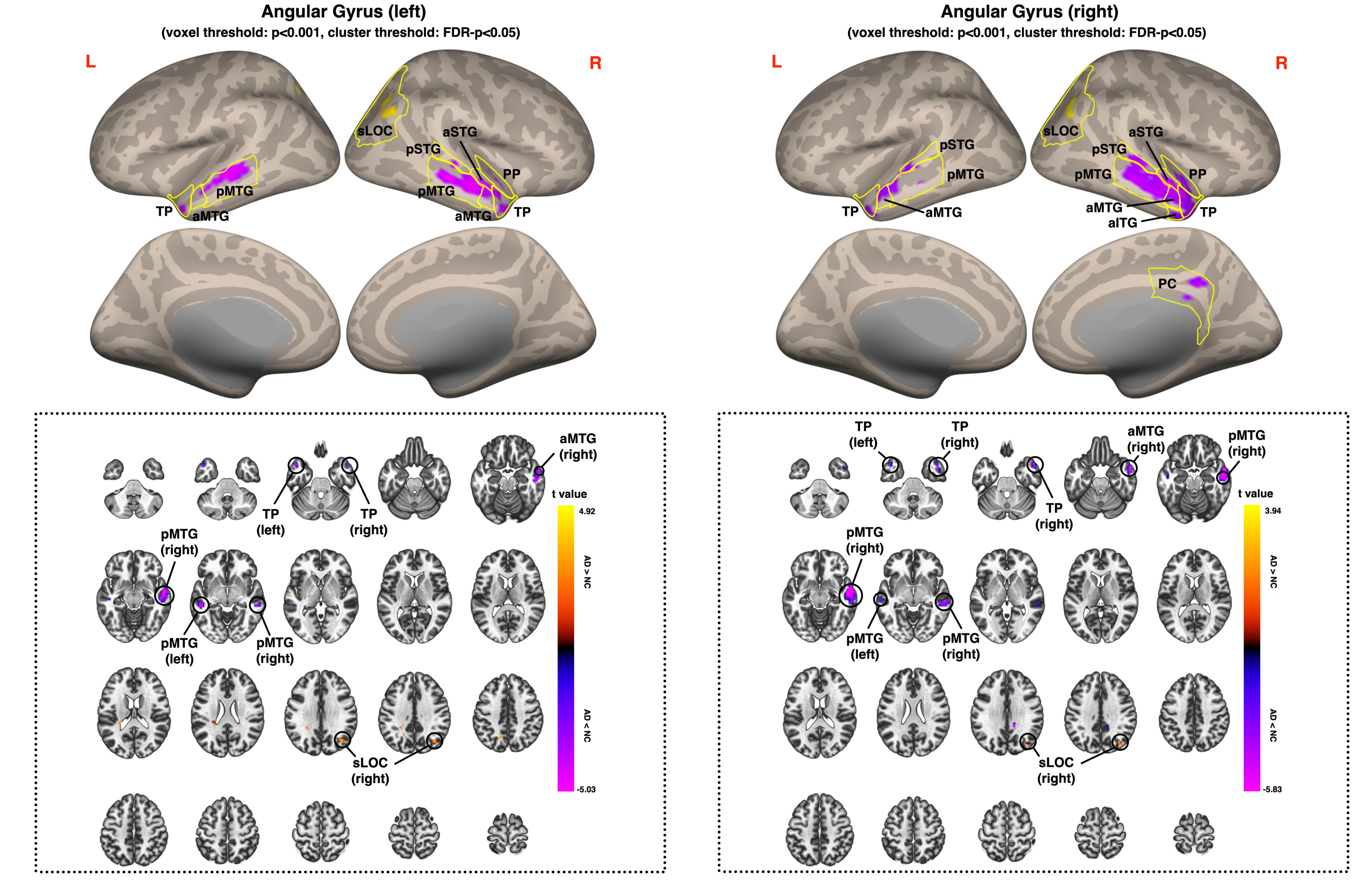}
	\caption[Voxel-level Connectivity Map of Left Angular Gyrus] {{\bf Voxel-level Connectivity Map of Left Angular Gyrus.} }
	\label{fig:fcmap.AG}
\end{figure}
%——————————————————————

\item n Figure \ref{fig:fcmap.pMTG}, the functional connections that are significantly decreased in the AD group with the left posterior Middle Temporal Gyrus (pMTG) are primarily located in the right posterior Middle Temporal Gyrus (pMTG), right Temporal Pole (TP), triangular part of right Inferior Frontal Gyrus (IFG tri), and right Paracingulate Gyrus (PaCiG).

\item Similarly, in Figure \ref{fig:fcmap.pMTG}, the functional connections that are significantly decreased in the AD group with the right posterior Middle Temporal Gyrus (pMTG) are primarily located in the left posterior Middle Temporal Gyrus (pMTG), left anterior Middle Temporal Gyrus (aMTG), bilateral Temporal Pole (TP), bilateral posterior Supramarginal Gyrus (pSMG), bilateral Angular Gyrus (AG), and right Parietal Operculum Cortex (PO).

% Figure: Voxel-level Connectivity Map of pMTG
%——————————————————————
\begin{figure}[]
	\centering
	\includegraphics[width=0.9\linewidth] {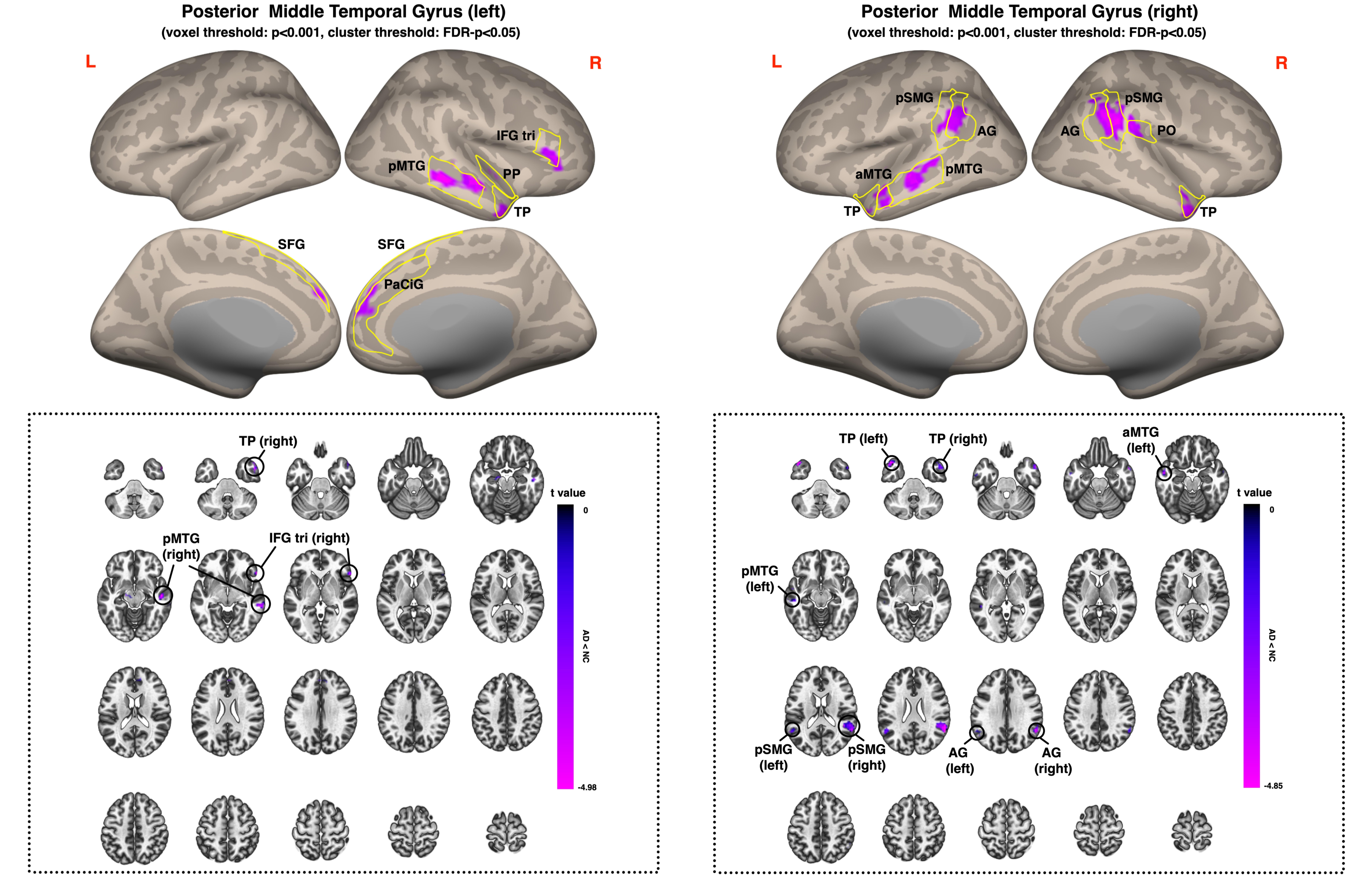}
	\caption[Voxel-level Connectivity Map of Left Posterior Middle Temporal Gyrus] {{\bf Voxel-level Connectivity Map of Left Posterior Middle Temporal Gyrus.} }
	\label{fig:fcmap.pMTG}
\end{figure}
%——————————————————————

\item In Figure \ref{fig:fcmap.TP}, the functional connections that are significantly decreased in the AD group with the left Temporal Pole (TP) are primarily located in the left posterior Supramarginal Gyrus (pSMG), left Angular Gyrus (AG), left posterior Middle Temporal Gyrus (pMTG), left temporooccipital Middle Temporal Gyrus (toMTG), left posterior Superior Temporal Gyrus (pSTG), left Planum Temporale (PT), Anterior Cingulate Gyrus (AC), Posterior Cingulate Gyrus (PC), Subcallosal Cortex (SubCalC), left Superior Frontal Gyrus (SFG), and left Paracingulate Gyrus (PaCiG). Similar results are observed in the right hemisphere.

\item Similarly, in Figure \ref{fig:fcmap.TP}, the functional connections that are significantly decreased in the AD group with the right Temporal Pole (TP) are primarily located in the right posterior Supramarginal Gyrus (pSMG), right anterior Supramarginal Gyrus (aSMG), right Postcentral Gyrus (PostCG), right Parietal Operculum Cortex (PO), right Central Opercular Cortex (CO), and right temporal gyrus (including the pMTG and pSTG). Likewise, decreased FCs are also observed in the left hemisphere. Note that increased FCs in the AD group are discovered in the right Posterior Cingulate Gyrus (PC), right Lingual Gyrus (LG), and the area 4 \& 5 of bilateral cerebellum (Cereb45). 

% Figure: Voxel-level Connectivity Map of TP
%——————————————————————
\begin{figure}[]
	\centering
	\includegraphics[width=0.9\linewidth]{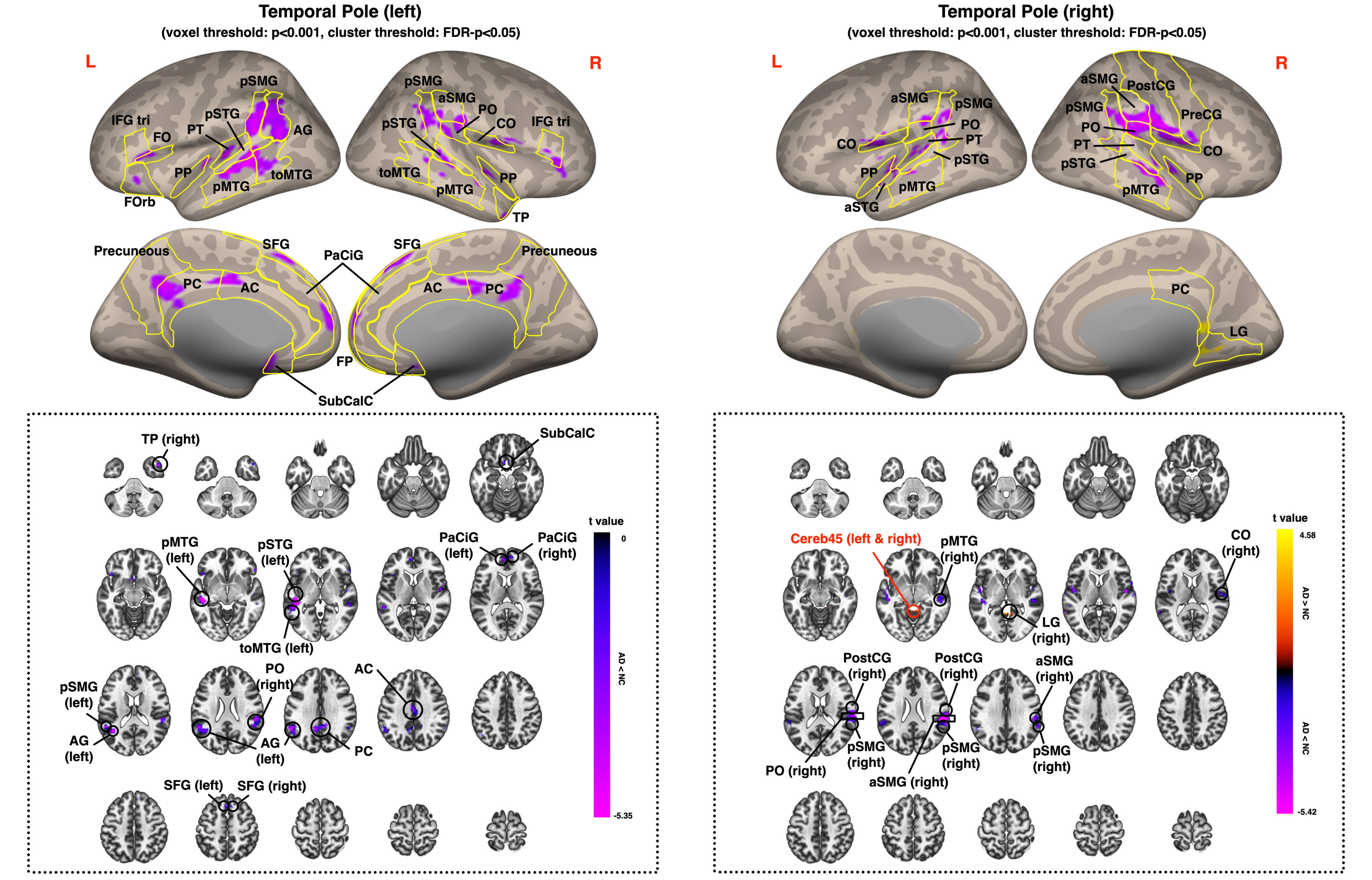}
	\caption[Voxel-level Connectivity Map of Left Temporal Pole] {{\bf Voxel-level Connectivity Map of Left Temporal Pole.} }
	\label{fig:fcmap.TP}
\end{figure}
%——————————————————————

\item In Figure \ref{fig:fcmap.IC}, the functional connections that are significantly decreased in the AD group with the left Insular Cortex (IC) are primarily located in bilateral Paracingulate Gyrus (PaCiG), Anterior Cingulate Gyrus (AC), Posterior Cingulate Gyrus (PC), right Precentral Gyrus (PreCG), right Precuneous, bilateral anterior Supramarginal Gyrus (aSMG), bilateral Parietal Operculum Cortex (PO), right Postcentral Gyrus (PostCG), right Precentral Gyrus (PreCG), and right Insular Cortex (IC).

\item Similarly, in Figure \ref{fig:fcmap.IC}, the functional connections that are significantly decreased in the AD group with the right Insular Cortex (IC) are primarily located in bilateral Paracingulate Gyrus (PaCiG), Anterior Cingulate Gyrus (AC), Posterior Cingulate Gyrus (PC), right Precentral Gyrus (PreCG), and right Precuneous.

% Figure: Voxel-level Connectivity Map of IC
%——————————————————————
\begin{figure}[]
	\centering
	\includegraphics[width=0.9\linewidth]{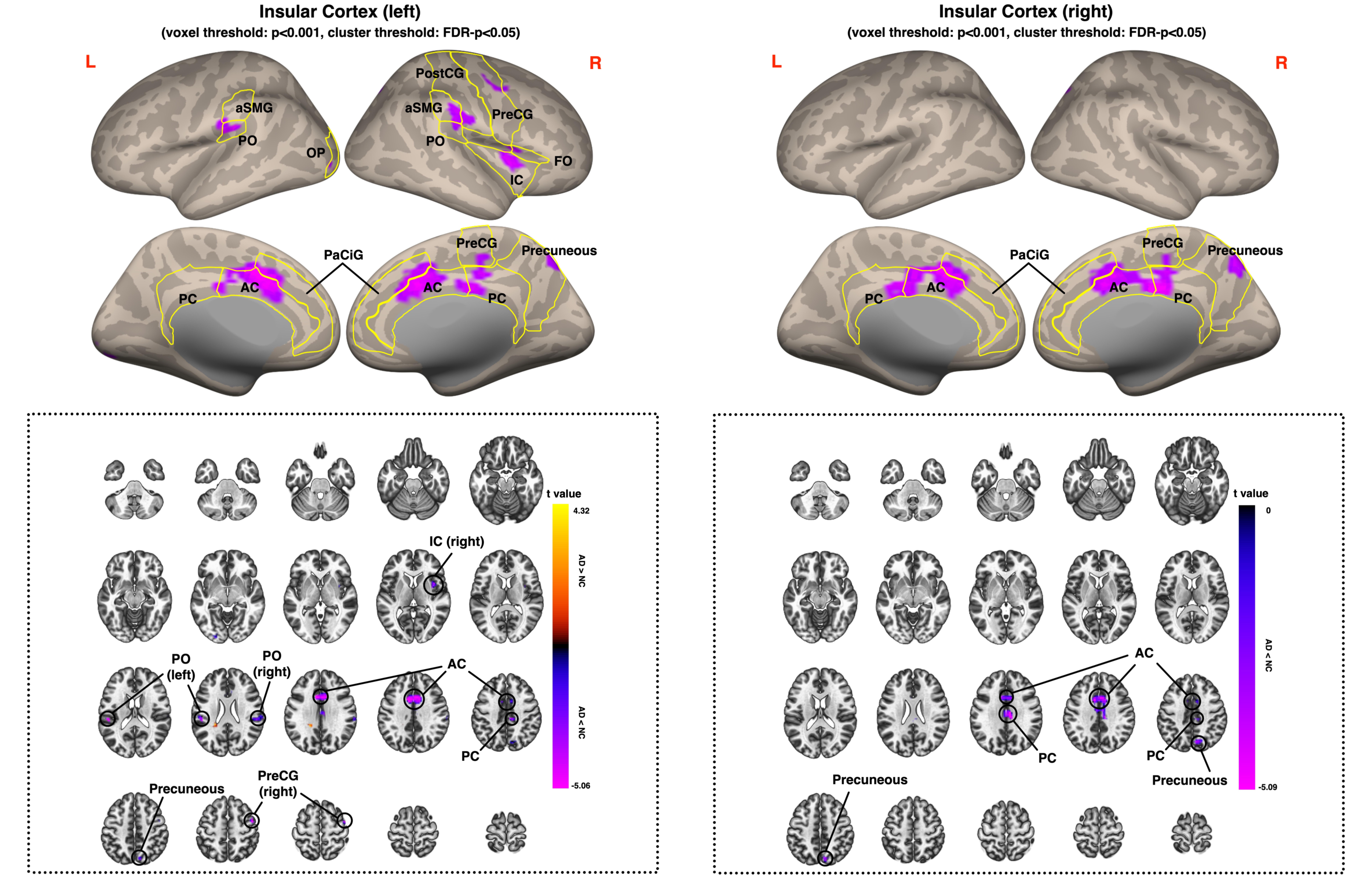}
	\caption[Voxel-level Connectivity Map of Left Insular Cortex] {{\bf Voxel-level Connectivity Map of Left Insular Cortex.} }
	\label{fig:fcmap.IC}
\end{figure}
%——————————————————————

\item In Figure \ref{fig:fcmap.PP}, the functional connections that are significantly decreased in the AD group with the left Planum Polare (PP) are primarily located in bilateral Precentral Gyrus (PreCG), bilateral Postcentral Gyrus (PostCG), bilateral anterior Supramarginal Gyrus (aSMG), bilateral Parietal Operculum Cortex (PO), bilateral Central Opercular Cortex (CO), right Insular Cortex (IC), bilateral anterior Superior Temporal Gyrus (aSTG), bilateral Planum Temporale (PT), Anterior Cingulate Gyrus (AC), Posterior Cingulate Gyrus (PC), bilateral Paracingulate Gyrus (PaCiG), and bilateral Supplementary Motor Cortex (SMA). Interestingly, the bilateral posterior Middle Temporal Gyrus (pMTG) show increased FCs with the left Planum Polare (PP).

\item  Similarly, in Figure \ref{fig:fcmap.PP}, the functional connections that are significantly decreased in the AD group with the right Planum Polare (PP) are primarily located in the Anterior Cingulate Gyrus (AC), bilateral Central Opercular Cortex (CO), left Parietal Operculum Cortex (PO), right Precentral Gyrus (PreCG), and right anterior Superior Temporal Gyrus (aSTG). Comparable to the left Planum Polare, increased functional connections with the right posterior Middle Temporal Gyrus (pMTG) are discovered. Increased FCs are also found in area 9 of the right cerebellum (Cereb9).

% Figure: Voxel-level Connectivity Map of PP
%——————————————————————
\begin{figure}[]
	\centering
	\includegraphics[width=0.9\linewidth]{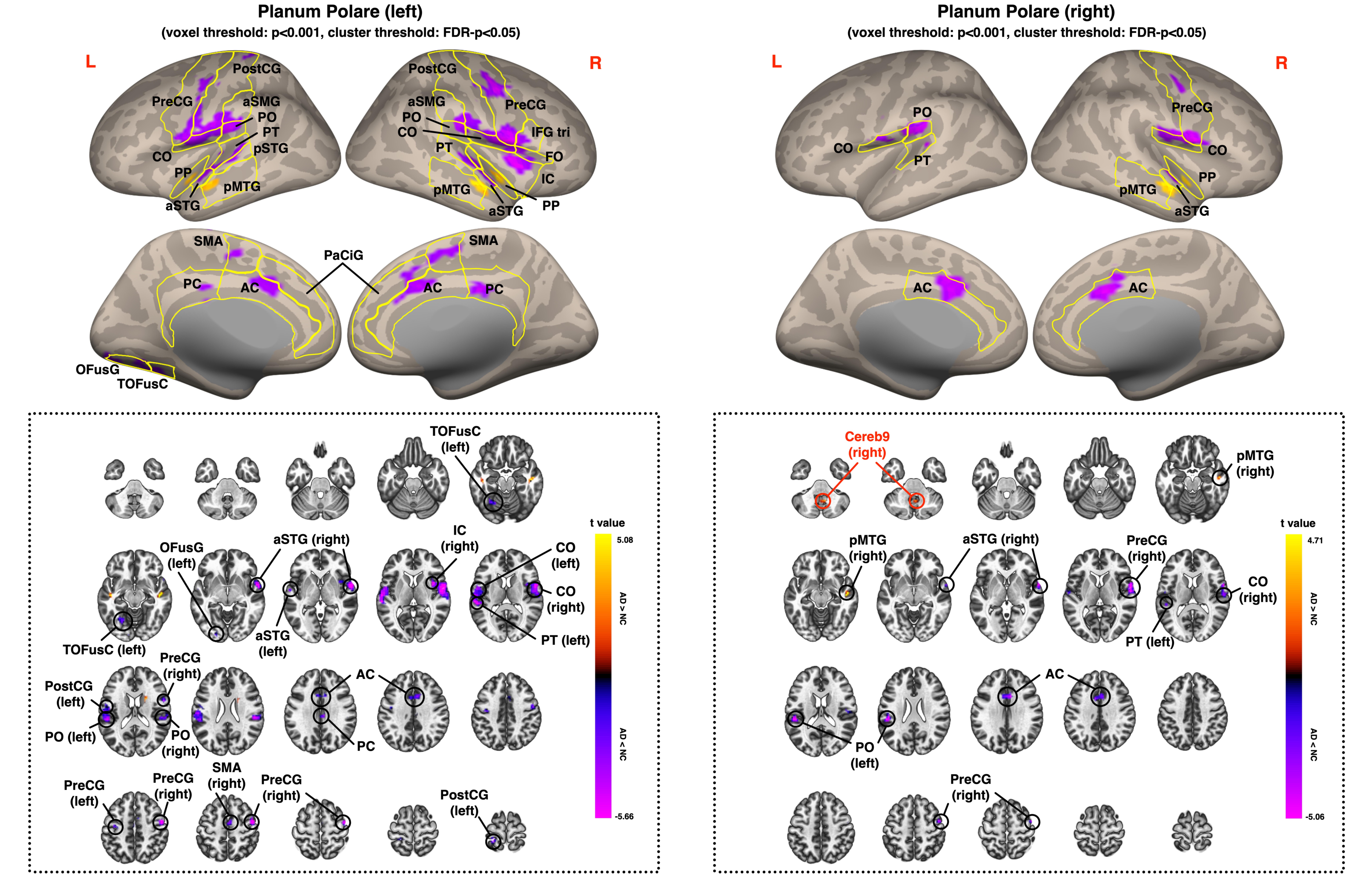}
	\caption[Voxel-level Connectivity Map of Left Planum Polare] {{\bf Voxel-level Connectivity Map of Left Planum Polare.} }
	\label{fig:fcmap.PP}
\end{figure}
%——————————————————————

\item In Figure \ref{fig:fcmap.PaCiG}, the functional connections that are significantly decreased in the AD group with the left Paracingulate Gyrus (PaCia) are primarily located in bilateral anterior Parahippocampal Gyrus (aPaHC), left posterior Parahippocampal Gyrus (pPaHC), Posterior Cingulate Gyrus (PC), left posterior Middle Temporal Gyrus (pMTG), and left anterior Middle Temporal Gyrus (aMTG). It is worth noting that decreased FCs are also observed in bilateral Hippocampus, which is consistent with the findings in the FC map of Hippocampus (Figure \ref{fig:fcmap.hippocampus}).

\item Similarly, in Figure \ref{fig:fcmap.PaCiG}, the functional connections that are significantly decreased in the AD group with the right Paracingulate Gyrus (PaCia) are primarily located in bilateral anterior Middle Temporal Gyrus (aMTG), bilateral posterior Middle Temporal Gyrus (pMTG), bilateral anterior Parahippocampal Gyrus (aPaHC), and left posterior Parahippocampal Gyrus (pPaHC). Comparable to the FC map of the left Paracingulate Gyrus, decreased FCs are also observed in bilateral Hippocampus.

% Figure: Voxel-level Connectivity Map of PaCiG
%——————————————————————
\begin{figure}[]
	\centering
	\includegraphics[width=0.9\linewidth]{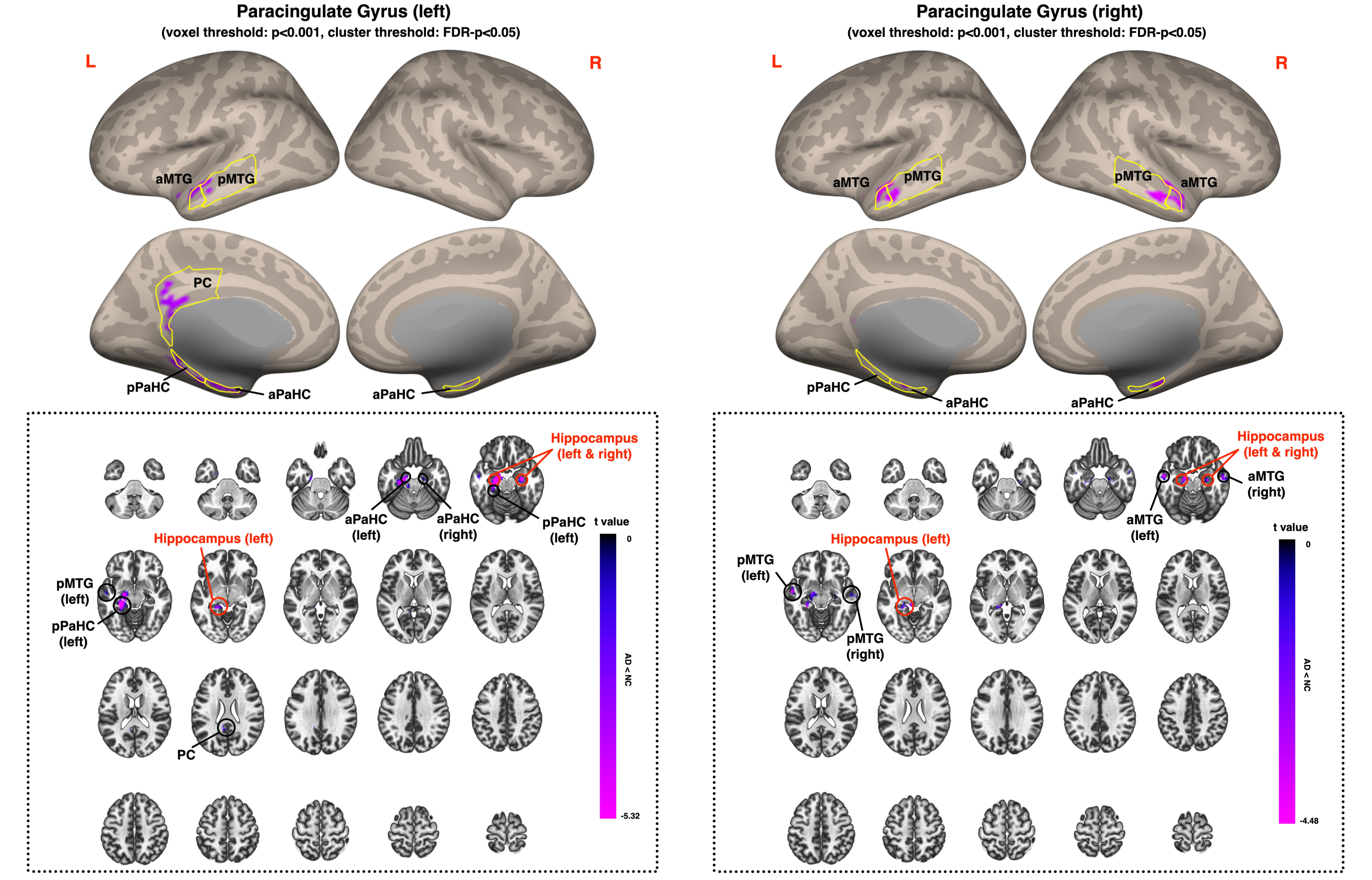}
	\caption[Voxel-level Connectivity Map of Left Paracingulate Gyrus] {{\bf Voxel-level Connectivity Map of Left Paracingulate Gyrus.} }
	\label{fig:fcmap.PaCiG}
\end{figure}
%——————————————————————

\item In Figure \ref{fig:fcmap.PO}, the functional connections that are significantly decreased in the AD group with the left Parietal Operculum Cortex (PO) are primarily located in the Anterior Cingulate Gyrus (AC) and Posterior Cingulate Gyrus (PC).

\item Similarly, in Figure \ref{fig:fcmap.PO}, the functional connections that are significantly decreased in the AD group with the right Parietal Operculum Cortex (PO) are primarily located in the Anterior Cingulate Gyrus (AC), Posterior Cingulate Gyrus (PC), bilateral Precentral Gyrus (PreCG), left Insular Cortex (IC), and left inferior Lateral Occipital Cortex (iLOC).

% Figure: Voxel-level Connectivity Map of PO
%——————————————————————
\begin{figure}[]
	\centering
	\includegraphics[width=0.9\linewidth]{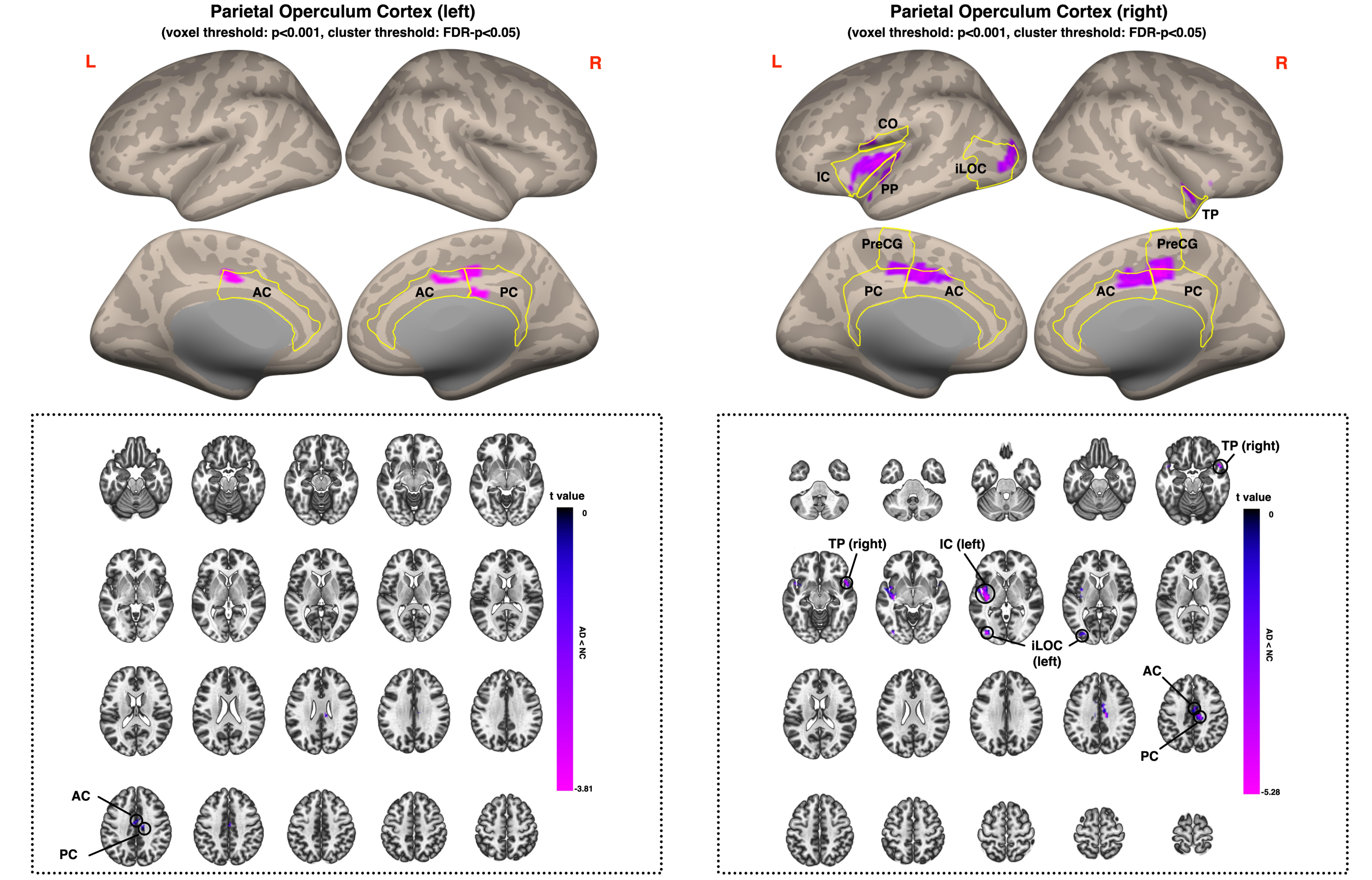}
	\caption[Voxel-level Connectivity Map of Left Parietal Operculum] {{\bf Voxel-level Connectivity Map of Left Parietal Operculum.} }
	\label{fig:fcmap.PO}
\end{figure}
%——————————————————————

\item In Figure \ref{fig:fcmap.FP}, the functional connections that are significantly decreased in the AD group with the left Frontal Pole (FP) are primarily located in bilateral Caudate.

\item Similarly, in Figure \ref{fig:fcmap.FP}, the functional connections that are significantly decreased in the AD group with the left Frontal Pole (FP) are mainly located in bilateral Caudate, right temporal gyrus (including the pMTG, aMTG, aITG, and TP), and Planum Polare (PP).

% Figure: Voxel-level Connectivity Map of FP
%——————————————————————
\begin{figure}[]
	\centering
	\includegraphics[width=0.9\linewidth]{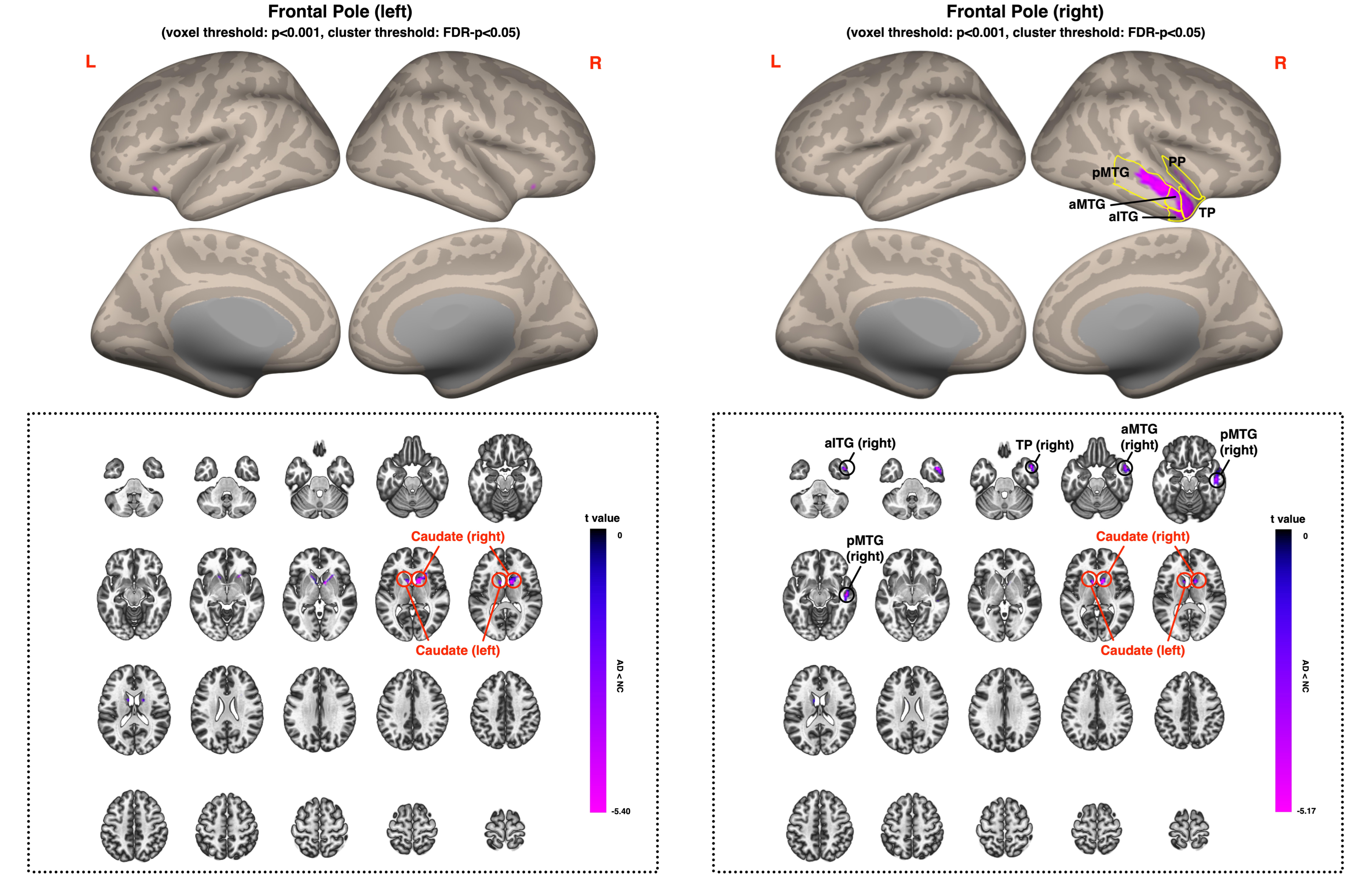}
	\caption[Voxel-level Connectivity Map of Left Frontal Pole] {{\bf Voxel-level Connectivity Map of Left Frontal Pole.} }
	\label{fig:fcmap.FP}
\end{figure}
%——————————————————————

\item In Figure \ref{fig:fcmap.MedFC}, the functional connections that are significantly decreased in the AD group with the Frontal Medial Cortex (MedFC) are primarily located in bilateral Hippocampus. Additionally, increased FCs with MedFC are discovered in bilateral Precuneous in the AD group.

% Figure: Voxel-level Connectivity Map of MedFC
%——————————————————————
\begin{figure}[]
	\centering
	\includegraphics[width=0.45\linewidth]{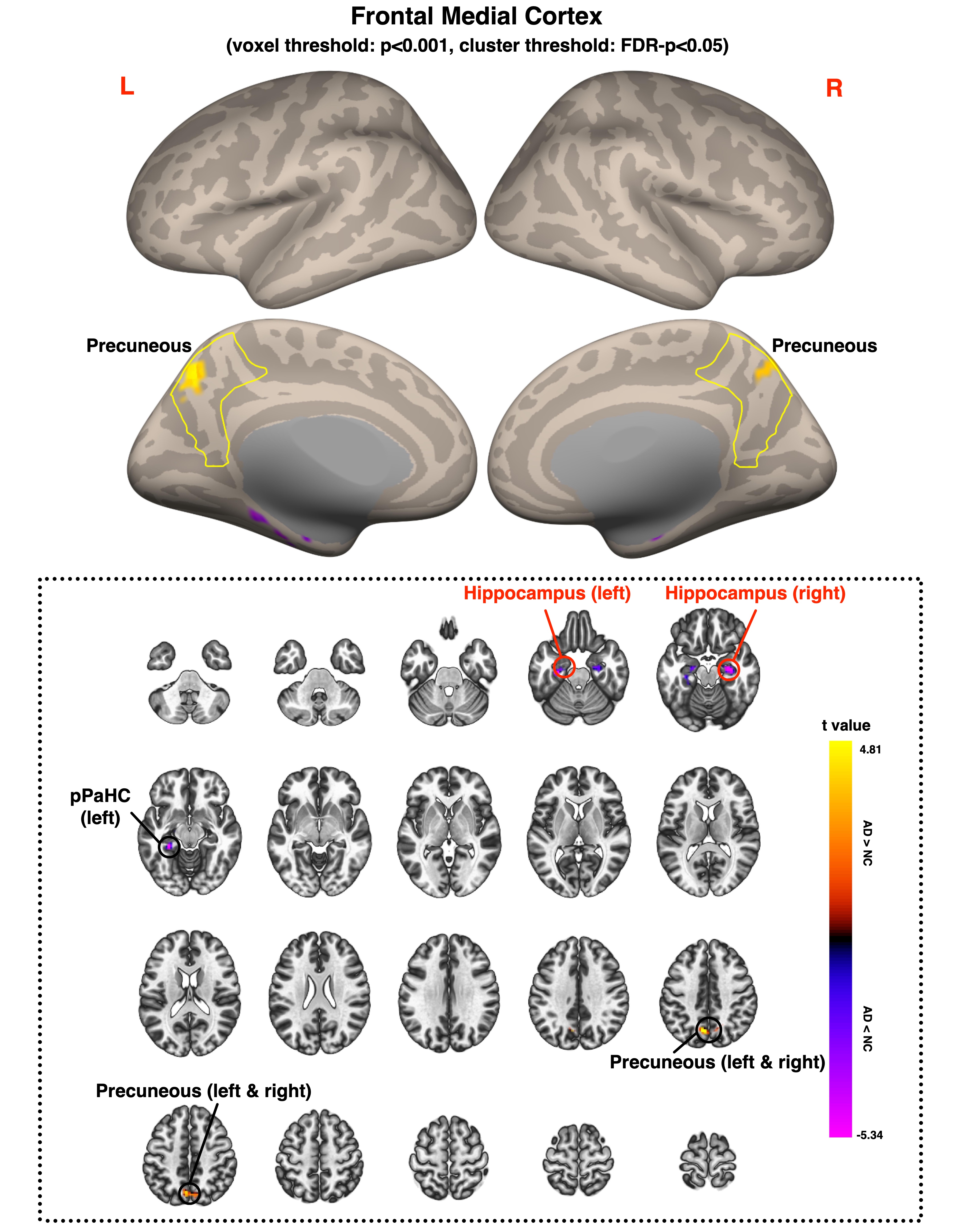}
	\caption[Voxel-level Connectivity Map of Frontal Medial Cortex] {{\bf Voxel-level Connectivity Map of Frontal Medial Cortex.} }
	\label{fig:fcmap.MedFC}
\end{figure}
%——————————————————————

\item In Figure \ref{fig:fcmap.Caudate}, the functional connections that are significantly decreased in the AD group with the left Caudate are primarily located in subcortical regions, including the left Amygdala, bilateral Putamen, left Pallidum, and bilateral Thalamus. Decreased FCs can also be found in the right superior Lateral Occipital Cortex (sLOC), right Angular Gyrus (AG), left Posterior Cingulate Gyrus (PC), and bilateral Frontal Pole (FP). Additionally, Figure \ref{fig:fcmap.Caudate} shows that the functional connections within the Caudate are significantly increased in the AD group compared with the NC group.

\item Similarly, in Figure \ref{fig:fcmap.Caudate}, the functional connections that are significantly decreased in the AD group with the right Caudate are primarily located in subcortical regions, including the bilateral Amygdala, left Putamen, left Pallidum, and bilateral Thalamus. Decreased FCs can also be found in the left Frontal Pole (FP) and left Posterior Cingulate Gyrus (PC). Additionally, Figure \ref{fig:fcmap.Caudate} shows that the functional connections within the Caudate are significantly increased in the AD group compared with the NC group. It is worth noting that increased FCs are observed in the Anterior Cingulate Gyrus (AC). This finding is consistent with the outcome of ROI-to-ROI analysis, where the FC between AC and the right Caudate is found to be significantly increased in the AD group (Figure \ref{fig:roi2roi.ring} (1), Table \ref{tab:result.roi2roi}).

% Figure: Voxel-level Connectivity Map of Caudate
%——————————————————————
\begin{figure}[]
	\centering
	\includegraphics[width=0.9\linewidth]{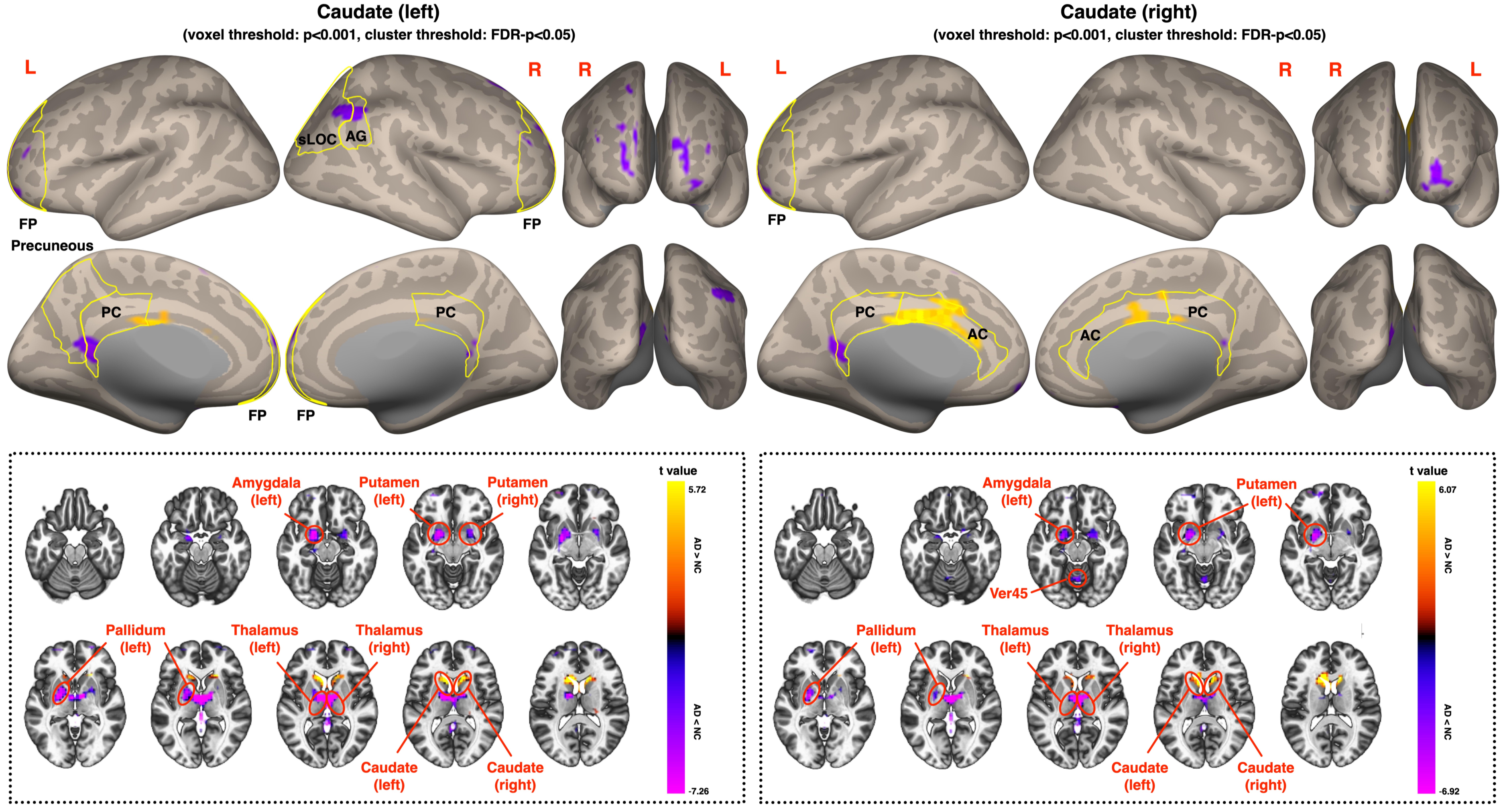}
	\caption[Voxel-level Connectivity Map of Left Caudate] {{\bf Voxel-level Connectivity Map of Left Caudate.} }
	\label{fig:fcmap.Caudate}
\end{figure}
%——————————————————————
\end{itemize}

\subsection{Discussion}
\subsubsection{Advantages and Disadvantages of ROI-to-ROI Method}
In ROI-to-ROI analyses, the mean signal of each ROI is computed prior to quantifying functional connections. As a result, one drawback of this approach is its strong dependence on the choice of brain atlas. Different brain atlases may lead to different conclusions. However, a benefit of this method is that it yields a distinct connectivity profile at the ROI level, making the interpretation of results straightforward.

\subsubsection{Advantages and Disadvantages of ROI-to-Voxel Method}
In the ROI-to-voxel method, solely the mean signal of the seed region is computed, thereby avoiding the issue of signal mixing. This approach allows for the generation of a comprehensive connectivity map for a given seed region. Nevertheless, a disadvantage of this method is the challenging interpretation of results due to the absence of an integrated connectivity profile. Specifically, the output of this analysis consists of numerous disconnected seed-to-voxel connectivity maps.

\subsubsection{Consistency of Altered ROI-to-ROI and ROI-to-Voxel Connectivity}
The consistency of findings from ROI-to-ROI and ROI-to-Voxel analyses are summarized as followed.

\begin{itemize}
\item In the AD group, a reduction in functional connections (FCs) between bilateral Hippocampus and bilateral Paracingulate Gyrus (PaCiG) is observed relative to the NC group. As can be seen in Figure \ref{fig:fcmap.hippocampus}, the voxels in the Paracingulate Gyrus are significantly and negatively correlated with the seed signal in the Hippocampus. This consistent finding is also evident in Figure \ref{fig:fcmap.PaCiG}, where significantly decreased FCs with the Paracingulate Gyrus are observed in both bilateral Hippocampus. These findings are in line with the results from ROI-to-ROI analyses, where decreased FCs between these regions are observed (Figure \ref{fig:roi2roi.ring} (1-4) and Table \ref{tab:result.roi2roi}).

\item The AD group exhibits a reduction in FCs between the Anterior Cingulate Gyrus (AC), Posterior Cingulate Gyrus (PC), and bilateral Hippocampus compared to the NC group (Figure \ref{fig:roi2roi.ring} (1) and Table \ref{tab:result.roi2roi}). Weakened FCs are observed in both ROI-to-ROI and ROI-to-Voxel analyses, including AC and left Hippocampus, PC and left Hippocampus (Figure \ref{fig:fcmap.hippocampus}), and PC and right Hippocampus (Figure \ref{fig:fcmap.hippocampus}).

\item Compared to the NC group, decreased FCs between bilateral Hippocampus and Frontal Medial Cortex (MedFC) are observed in the AD group (Figure \ref{fig:roi2roi.ring} (1-3), Table \ref{tab:result.roi2roi}, and Figure \ref{fig:fcmap.hippocampus}). Consistent findings are also evident in Figure \ref{fig:fcmap.MedFC}

\item In the AD group, a decrease in FC between the left and right posterior Middle Temporal Gyrus (pMTG) is observed in two analyses (Figure \ref{fig:roi2roi.ring} (2), Table \ref{tab:result.roi2roi}, and Figure \ref{fig:fcmap.pMTG}) relative to the NC group.

\item Compared to the NC group, a reduction in FC between the right posterior Middle Temporal Gyrus (pMTG) and right Angular Gyrus (AG) (Figure \ref{fig:roi2roi.ring} (1-3), Table \ref{tab:result.roi2roi}, and Figure \ref{fig:fcmap.AG} \& \ref{fig:fcmap.pMTG}) is observed in the AD group. Similarly, decreased FC between the right pMTG and left AG (Figure \ref{fig:roi2roi.ring} (1), Table \ref{tab:result.roi2roi}, and Figure \ref{fig:fcmap.AG} \& \ref{fig:fcmap.pMTG}) is also observed in the AD group. These results are verified by two analyses in this study. However, some findings in the ROI-to-Voxel analysis are not statistically significant in the ROI-to-ROI analysis. For example, the FC between the left AG and left pMTG (Figure \ref{fig:fcmap.AG} \& \ref{fig:fcmap.pMTG}) as well as the FC between the right AG and left pMTG (Figure \ref{fig:fcmap.AG} \& \ref{fig:fcmap.pMTG}) are significantly decreased in the AD group in the ROI-to-Voxel analysis, yet not statistically significant in the ROI-to-ROI analysis.

\item In the AD group, a decrease in FC between the left anterior Parahippocampal Gyrus (aPaHC) and left Paracingulate Gyrus (PaCiG) is observed in both the ROI-to-ROI and ROI-to-Voxel analyses (Figure \ref{fig:roi2roi.ring} (1, 2), Table \ref{tab:result.roi2roi}, and Figure \ref{fig:fcmap.aPaHC} \& \ref{fig:fcmap.PaCiG}). Additionally, the following FCs are significantly weakened in the AD group in the ROI-to-Voxel analysis: left PaCiG \& right aPaHC (Figure \ref{fig:fcmap.PaCiG} \& \ref{fig:fcmap.aPaHC}), right PaCiG \& right aPaHC (Figure \ref{fig:fcmap.PaCiG} \& \ref{fig:fcmap.aPaHC}), and right PaCiG \& left aPaHC (Figure \ref{fig:fcmap.PaCiG} \& \ref{fig:fcmap.aPaHC}).
\item In the AD group, a decrease in FC between the right Temporal Pole (TP) and right Parietal Operculum (PO) is observed in two analyses (Figure \ref{fig:roi2roi.ring} (1-3), Table \ref{tab:result.roi2roi}, and Figure \ref{fig:fcmap.TP} \& \ref{fig:fcmap.PO}). Additionally, the FC between the left TP and right PO is significantly decreased in the AD group (Figure \ref{fig:roi2roi.ring} (1), Table \ref{tab:result.roi2roi}, and Figure \ref{fig:fcmap.TP}).

\item In the AD group, a decrease in FC between the left Frontal Orbital Cortex (FOrb) and left Temporal Pole (TP) is observed in two analyses (Figure \ref{fig:roi2roi.ring} (1, 2), Table \ref{tab:result.roi2roi}, and Figure \ref{fig:fcmap.TP}).

\item In the AD group, a decrease in FCs between bilateral Frontal Pole (FP) and bilateral Caudate (Table \ref{tab:result.roi2roi}) is observed in two analyses: left FP \& left Caudate (Figure \ref{fig:roi2roi.ring} (1), and Figure \ref{fig:fcmap.FP} \& \ref{fig:fcmap.Caudate}); left FP \& right Caudate (Figure \ref{fig:roi2roi.ring} (1), and Figure \ref{fig:fcmap.FP} \& \ref{fig:fcmap.Caudate}); right FP \& right Caudate (Figure \ref{fig:roi2roi.ring} (1), and Figure \ref{fig:fcmap.FP} \& \ref{fig:fcmap.Caudate}); right FP \& left Caudate (Figure \ref{fig:roi2roi.ring} (1), and Figure \ref{fig:fcmap.FP} \& \ref{fig:fcmap.Caudate}).

\item In the AD group, a decrease in FC between the right Caudate and Ver45 is observed in two analyses (Figure \ref{fig:roi2roi.ring} (1-3) \& Figure \ref{fig:fcmap.Caudate}).

\item In the AD group, a decrease in FCs between bilateral Caudate and Putamen is observed in two analyses: left Caudate \& left Putamen, left Caudate \& right Putamen, and right Caudate \& left Putamen (Figure \ref{fig:roi2roi.ring} (1), Table \ref{tab:result.roi2roi}, and Figure \ref{fig:fcmap.Caudate}).

\item In the AD group, a decrease in FC between the left Insular Cortex (IC) and right Parietal Operculum (PO) is observed in two analyses (Figure \ref{fig:roi2roi.ring} (1, 2), Table \ref{tab:result.roi2roi}, and Figure \ref{fig:fcmap.IC} \& \ref{fig:fcmap.PO}).

\item In the AD group, the decreased FCs between bilateral Planum Polare (PP) and Parietal Operculum (PO) (Figure \ref{fig:roi2roi.ring} (1) and Table \ref{tab:result.roi2roi}) are highly consistent in both the ROI-to-ROI and ROI-to-Voxel analyses. These FCs are left PP \& left PO (Figure \ref{fig:fcmap.PP}), left PP \& right PO (Figure \ref{fig:fcmap.PP} \& \ref{fig:fcmap.PO}), and right PP \& left PO (Figure \ref{fig:fcmap.PP}).

\item In the AD group, the decreased FCs between bilateral Planum Polare (PP) and Central Opercular Cortex (CO) (Figure \ref{fig:roi2roi.ring} (1) and Table \ref{tab:result.roi2roi}) are highly consistent in both the ROI-to-ROI and ROI-to-Voxel analyses. These FCs are left PP \& left CO, left PP \& right CO (Figure \ref{fig:fcmap.PP}), and right PP \& right CO (Figure \ref{fig:fcmap.PP}).

\end{itemize}

\section{Conclusions}
\begin{figure}[]
	\centering
	\includegraphics[width=0.6\linewidth] {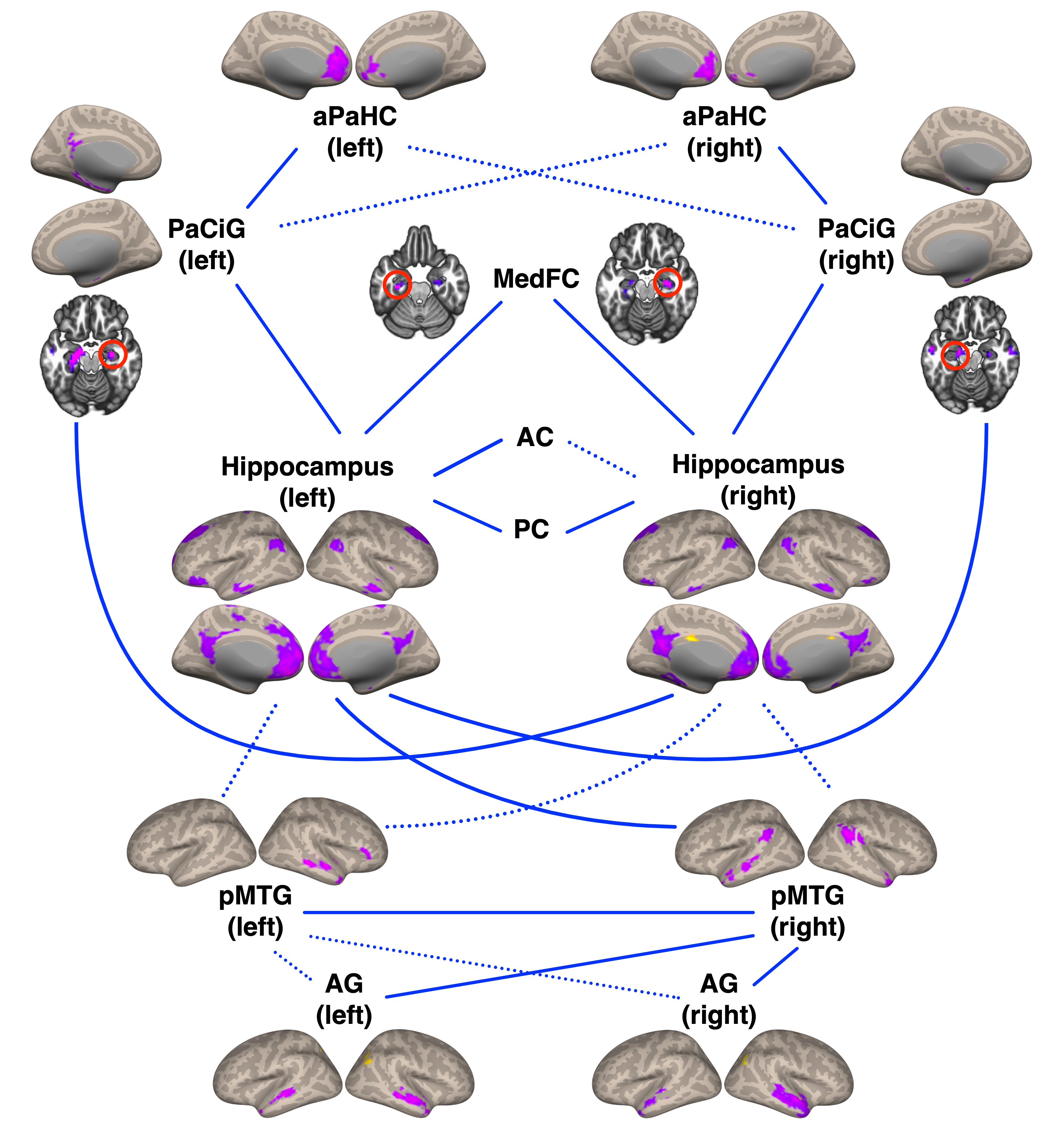}
	\caption[Decreased Functional Connections in AD Group (circuit 1)] {{\bf Decreased Functional Connections in AD Group (circuit 1).} Solid line: results observed in both ROI-to-ROI and ROI-to-Voxel analyses. Dotted line: results observed from only one of the analyses.}
	\label{fig:circuit_1}
\end{figure}

\begin{figure}[]
	\centering
	\includegraphics[width=0.6\linewidth] {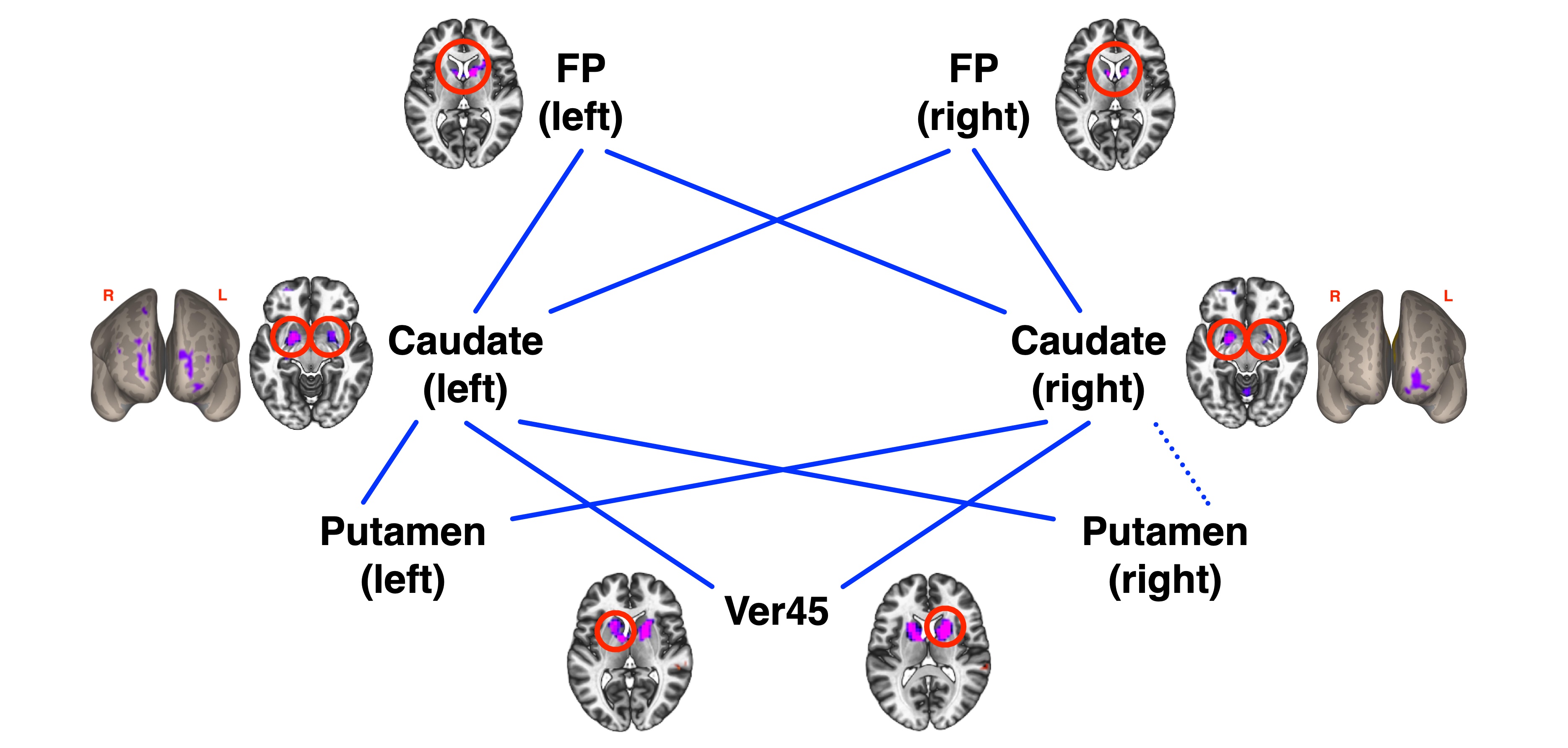}
	\caption[Decreased Functional Connections in AD Group (circuit 2)] {{\bf Decreased Functional Connections in AD Group (circuit 2).} Solid line: results observed in both ROI-to-ROI and ROI-to-Voxel analyses. Dotted line: results observed from only one of the analyses.}
	\label{fig:circuit_2}
\end{figure}

\begin{figure}[]
	\centering
	\includegraphics[width=0.6\linewidth] {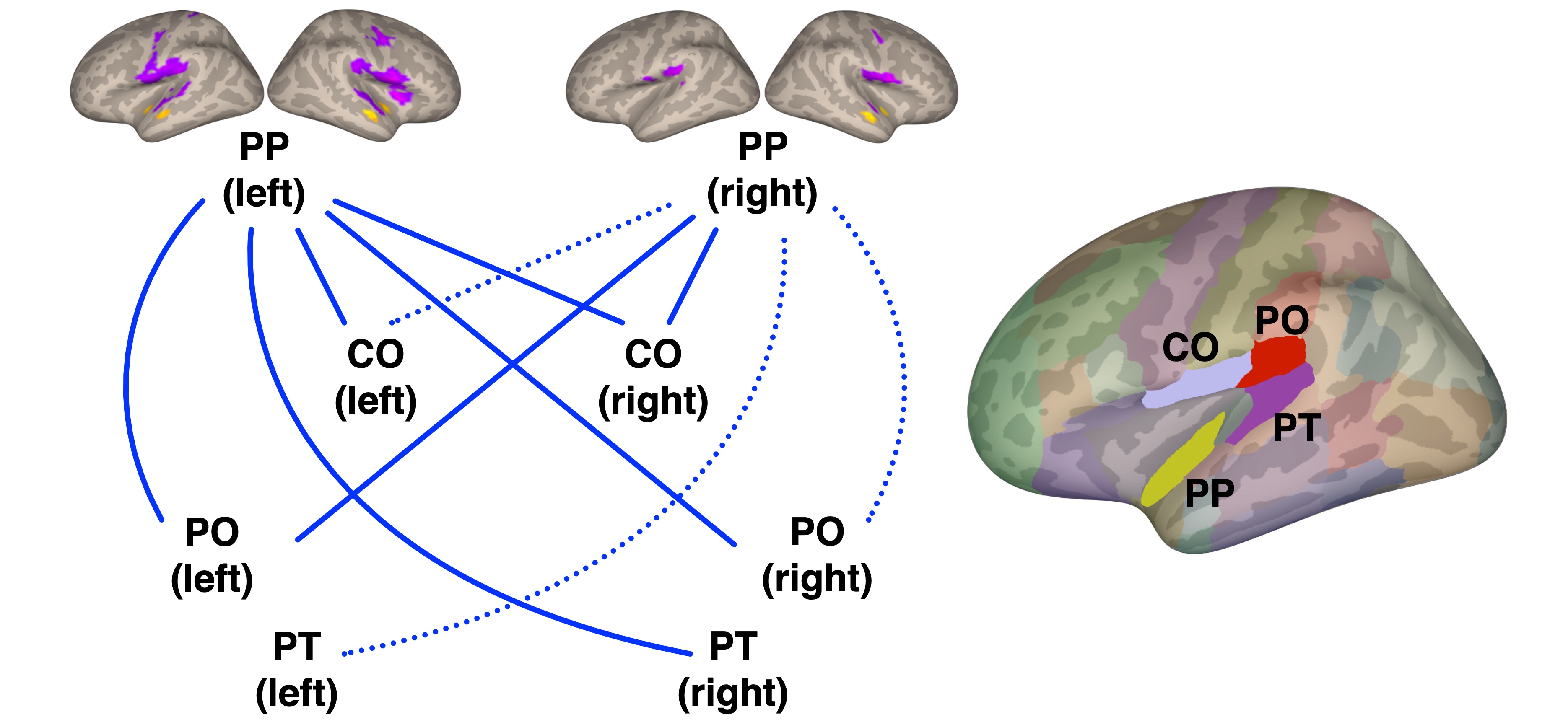}
	\caption[Decreased Functional Connections in AD Group (circuit 3)] {{\bf Decreased Functional Connections in AD Group (circuit 3).} Solid line: results observed in both ROI-to-ROI and ROI-to-Voxel analyses. Dotted line: results observed from only one of the analyses.}
	\label{fig:circuit_3}
\end{figure}

From the ROI-to-ROI and ROI-to-Voxel functional connectivity analyses, we draw the following conclusions:

\begin{enumerate}
\item The AD group presents a significant reduction in functional connections when compared to the NC group. Specifically, the following FCs demonstrate decreased connectivity: bilateral Hippocampus and bilateral PaCiG, bilateral Hippocampus and MedFC, bilateral aPaHC and bilateral PaCiG, bilateral Hippocampus and PC, bilateral Hippocampus and AC, left pMTG and right pMTG, bilateral AG and bilateral pMTG, bilateral FP and bilateral Caudate, bilateral Caudate and bilateral Putamen, bilateral Caudate and Ver45, bilateral PP and bilateral CO, bilateral PP and bilateral PO, and left PP and right PT. These results are depicted in Figure \ref{fig:circuit_1}, Figure \ref{fig:circuit_2}, and Figure \ref{fig:circuit_3}.

\item The functional connectivity of the brain network in AD patients is extensively decreased, particularly in regions such as bilateral Hippocampus, MedFC, bilateral PaCiG, right pMTG, right AG, bilateral Caudate, left Putamen, Ver45, right PO, and right TP, as well as bilateral FP, which exhibit a marked decrease in FCs with a strong effect size.
\end{enumerate}

% Figure: circuit_1
%——————————————————————
%——————————————————————

% Figure: circuit_2
%——————————————————————
%——————————————————————

% Figure: circuit_3
%——————————————————————
%——————————————————————

%Bibliography
\newpage
\bibliographystyle{unsrt}  
\bibliography{references}

\begin{thebibliography}{10}

\bibitem{Blennow_2006}
Kaj Blennow, Mony~J de~Leon, and Henrik Zetterberg.
\newblock Alzheimer's disease.
\newblock {\em The Lancet}, 368(9533):387--403, Jul 2006.

\bibitem{hardy1991amyloid}
John Hardy and David Allsop.
\newblock Amyloid deposition as the central event in the aetiology of
  alzheimer's disease.
\newblock {\em Trends in pharmacological sciences}, 12:383--388, 1991.

\bibitem{iqbal2005tau}
Khalid Iqbal, Alejandra del~C Alonso, She Chen, M~Omar Chohan, Ezzat El-Akkad,
  Cheng-Xin Gong, Sabiha Khatoon, Bin Li, Fei Liu, Abdur Rahman, et~al.
\newblock Tau pathology in alzheimer disease and other tauopathies.
\newblock {\em Biochimica et Biophysica Acta (BBA)-Molecular Basis of Disease},
  1739(2-3):198--210, 2005.

\bibitem{pariente2005alzheimer}
J{\'e}r{\'e}mie Pariente, Susanna Cole, Richard Henson, Linda Clare, Angus
  Kennedy, Martin Rossor, Lisa Cipoloti, Mich{\`e}le Puel, Jean~Francois
  Demonet, Francois Chollet, et~al.
\newblock Alzheimer's patients engage an alternative network during a memory
  task.
\newblock {\em Annals of Neurology: Official Journal of the American
  Neurological Association and the Child Neurology Society}, 58(6):870--879,
  2005.

\bibitem{celone2006alterations}
Kim~A Celone, Vince~D Calhoun, Bradford~C Dickerson, Alireza Atri, Elizabeth~F
  Chua, Saul~L Miller, Kristina DePeau, Doreen~M Rentz, Dennis~J Selkoe,
  Deborah Blacker, et~al.
\newblock Alterations in memory networks in mild cognitive impairment and
  alzheimer's disease: an independent component analysis.
\newblock {\em Journal of Neuroscience}, 26(40):10222--10231, 2006.

\bibitem{biswal1995functional}
Bharat Biswal, F~Zerrin~Yetkin, Victor~M Haughton, and James~S Hyde.
\newblock Functional connectivity in the motor cortex of resting human brain
  using echo-planar mri.
\newblock {\em Magnetic resonance in medicine}, 34(4):537--541, 1995.

\bibitem{chen2011classification}
Gang Chen, B~Douglas Ward, Chunming Xie, Wenjun Li, Zhilin Wu, Jennifer~L
  Jones, Malgorzata Franczak, Piero Antuono, and Shi-Jiang Li.
\newblock Classification of alzheimer disease, mild cognitive impairment, and
  normal cognitive status with large-scale network analysis based on
  resting-state functional mr imaging.
\newblock {\em Radiology}, 259(1):213--221, 2011.

\bibitem{wang2006changes}
Liang Wang, Yufeng Zang, Yong He, Meng Liang, Xinqing Zhang, Lixia Tian, Tao
  Wu, Tianzi Jiang, and Kuncheng Li.
\newblock Changes in hippocampal connectivity in the early stages of
  alzheimer's disease: evidence from resting state fmri.
\newblock {\em Neuroimage}, 31(2):496--504, 2006.

\bibitem{agosta2012resting}
Federica Agosta, Michela Pievani, Cristina Geroldi, Massimiliano Copetti,
  Giovanni~B Frisoni, and Massimo Filippi.
\newblock Resting state fmri in alzheimer's disease: beyond the default mode
  network.
\newblock {\em Neurobiology of aging}, 33(8):1564--1578, 2012.

\bibitem{binnewijzend2012resting}
Maja~AA Binnewijzend, Menno~M Schoonheim, Ernesto Sanz-Arigita, Alle~Meije
  Wink, Wiesje~M van~der Flier, Nelleke Tolboom, Sofie~M Adriaanse, Jessica~S
  Damoiseaux, Philip Scheltens, Bart~NM van Berckel, et~al.
\newblock Resting-state fmri changes in alzheimer's disease and mild cognitive
  impairment.
\newblock {\em Neurobiology of aging}, 33(9):2018--2028, 2012.

\bibitem{koch2012diagnostic}
Walter Koch, Stephan Teipel, Sophia Mueller, Jens Benninghoff, Maxmilian
  Wagner, Arun~LW Bokde, Harald Hampel, Ute Coates, Maximilian Reiser, and
  Thomas Meindl.
\newblock Diagnostic power of default mode network resting state fmri in the
  detection of alzheimer's disease.
\newblock {\em Neurobiology of aging}, 33(3):466--478, 2012.

\bibitem{honey2009predicting}
CJ~Honey, O~Sporns, Leila Cammoun, Xavier Gigandet, Jean-Philippe Thiran, Reto
  Meuli, and Patric Hagmann.
\newblock Predicting human resting-state functional connectivity from
  structural connectivity.
\newblock {\em Proceedings of the National Academy of Sciences},
  106(6):2035--2040, 2009.

\bibitem{wang2015understanding}
Zhijiang Wang, Zhengjia Dai, Gaolang Gong, Changsong Zhou, and Yong He.
\newblock Understanding structural-functional relationships in the human brain:
  a large-scale network perspective.
\newblock {\em The Neuroscientist}, 21(3):290--305, 2015.

\bibitem{mivsic2016network}
Bratislav Mi{\v{s}}i{\'c}, Richard~F Betzel, Marcel~A De~Reus, Martijn~P Van
  Den~Heuvel, Marc~G Berman, Anthony~R McIntosh, and Olaf Sporns.
\newblock Network-level structure-function relationships in human neocortex.
\newblock {\em Cerebral Cortex}, 26(7):3285--3296, 2016.

\bibitem{achard2006resilient}
Sophie Achard, Raymond Salvador, Brandon Whitcher, John Suckling, and
  ED~Bullmore.
\newblock A resilient, low-frequency, small-world human brain functional
  network with highly connected association cortical hubs.
\newblock {\em Journal of Neuroscience}, 26(1):63--72, 2006.

\bibitem{hagmann2008mapping}
Patric Hagmann, Leila Cammoun, Xavier Gigandet, Reto Meuli, Christopher~J
  Honey, Van~J Wedeen, and Olaf Sporns.
\newblock Mapping the structural core of human cerebral cortex.
\newblock {\em PLoS biology}, 6(7), 2008.

\bibitem{buckner2009cortical}
Randy~L Buckner, Jorge Sepulcre, Tanveer Talukdar, Fenna~M Krienen, Hesheng
  Liu, Trey Hedden, Jessica~R Andrews-Hanna, Reisa~A Sperling, and Keith~A
  Johnson.
\newblock Cortical hubs revealed by intrinsic functional connectivity: mapping,
  assessment of stability, and relation to alzheimer's disease.
\newblock {\em Journal of neuroscience}, 29(6):1860--1873, 2009.

\bibitem{sporns2007identification}
Olaf Sporns, Christopher~J Honey, and Rolf K{\"o}tter.
\newblock Identification and classification of hubs in brain networks.
\newblock {\em PloS one}, 2(10), 2007.

\bibitem{power2013evidence}
Jonathan~D Power, Bradley~L Schlaggar, Christina~N Lessov-Schlaggar, and
  Steven~E Petersen.
\newblock Evidence for hubs in human functional brain networks.
\newblock {\em Neuron}, 79(4):798--813, 2013.

\bibitem{thomas2011organization}
BT~Thomas~Yeo, Fenna~M Krienen, Jorge Sepulcre, Mert~R Sabuncu, Danial
  Lashkari, Marisa Hollinshead, Joshua~L Roffman, Jordan~W Smoller, Lilla
  Z{\"o}llei, Jonathan~R Polimeni, et~al.
\newblock The organization of the human cerebral cortex estimated by intrinsic
  functional connectivity.
\newblock {\em Journal of neurophysiology}, 106(3):1125--1165, 2011.

\bibitem{power2011functional}
Jonathan~D Power, Alexander~L Cohen, Steven~M Nelson, Gagan~S Wig, Kelly~Anne
  Barnes, Jessica~A Church, Alecia~C Vogel, Timothy~O Laumann, Fran~M Miezin,
  Bradley~L Schlaggar, et~al.
\newblock Functional network organization of the human brain.
\newblock {\em Neuron}, 72(4):665--678, 2011.

\bibitem{buckner2011organization}
Randy~L Buckner, Fenna~M Krienen, Angela Castellanos, Julio~C Diaz, and
  BT~Thomas Yeo.
\newblock The organization of the human cerebellum estimated by intrinsic
  functional connectivity.
\newblock {\em Journal of neurophysiology}, 106(5):2322--2345, 2011.

\bibitem{wu2016distinct}
Yan Wu, Yaqin Zhang, Yong Liu, Jieqiong Liu, Yunyun Duan, Xuehu Wei, Junjie
  Zhuo, Kuncheng Li, Xinqin Zhang, Chunshui Yu, et~al.
\newblock Distinct changes in functional connectivity in posteromedial cortex
  subregions during the progress of alzheimer's disease.
\newblock {\em Frontiers in neuroanatomy}, 10:41, 2016.

\bibitem{zheng2017altered}
Weimin Zheng, Xingyun Liu, Haiqing Song, Kuncheng Li, and Zhiqun Wang.
\newblock Altered functional connectivity of cognitive-related cerebellar
  subregions in alzheimer's disease.
\newblock {\em Frontiers in aging neuroscience}, 9:143, 2017.

\bibitem{lamontagne2018oasis}
Pamela~J LaMontagne, Sarah Keefe, Wallace Lauren, Chengjie Xiong, Elizabeth~A
  Grant, Krista~L Moulder, John~C Morris, Tammie~LS Benzinger, and Daniel~S
  Marcus.
\newblock Oasis-3: Longitudinal neuroimaging, clinical, and cognitive dataset
  for normal aging and alzheimer's disease.
\newblock {\em Alzheimer's \& Dementia: The Journal of the Alzheimer's
  Association}, 14(7):P1097, 2018.

\bibitem{yao2023altered}
Yongcheng Yao.
\newblock Altered topological properties of functional brain network associated
  with alzheimer's disease, 2023.

\bibitem{whitfield2012conn}
Susan Whitfield-Gabrieli and Alfonso Nieto-Castanon.
\newblock Conn: a functional connectivity toolbox for correlated and
  anticorrelated brain networks.
\newblock {\em Brain connectivity}, 2(3):125--141, 2012.

\bibitem{power2015recent}
Jonathan~D Power, Bradley~L Schlaggar, and Steven~E Petersen.
\newblock Recent progress and outstanding issues in motion correction in
  resting state fmri.
\newblock {\em Neuroimage}, 105:536--551, 2015.

\bibitem{behzadi2007component}
Yashar Behzadi, Khaled Restom, Joy Liau, and Thomas~T Liu.
\newblock A component based noise correction method (compcor) for bold and
  perfusion based fmri.
\newblock {\em Neuroimage}, 37(1):90--101, 2007.

\end{thebibliography}

\end{document}